\documentclass[11pt]{iopart}

\usepackage[british]{babel}
\usepackage{color}
\usepackage{mathrsfs}
\usepackage{setstack}
\usepackage{iopams}
\usepackage[T1]{fontenc}
\usepackage{graphicx}
\usepackage{cancel}

\begin{document}

\title[Crossover from anomalous to normal diffusion]{Crossover from anomalous to normal
diffusion: truncated power-law noise correlations and applications to dynamics
in lipid bilayers}

\author{Daniel Molina-Garcia$^{\dagger,\sharp,\$}$, Trifce Sandev$^{\flat,
\ddagger,\S}$, Hadiseh Safdari$^{\pounds}$, Gianni Pagnini$^{\dagger,\&}$,
Aleksei Chechkin$^{\sharp,\P}$, and Ralf Metzler$^{\sharp}$}
\address{$\dagger$ BCAM - Basque Center for Applied Mathematics, Alameda de
Mazarredo 14, E-48009 Bilbao, Basque Country, Spain\\
$\sharp$ Institute of Physics \& Astronomy, University of Potsdam, D-14776
Potsdam-Golm, Germany\\
$\$$ University of the Basque Country UPV/EHU, Barrio Sarriena s/n, 48940 Leioa,
Basque Country, Spain\\
$\flat$ Radiation Safety Directorate, Partizanski odredi 143, P.O. Box 22, 1020
Skopje, Macedonia\\
$\ddagger$ Institute of Physics, Faculty of Natural Sciences and Mathematics, Ss
Cyril and Methodius University, P.O. Box 162, 1001 Skopje, Macedonia\\
$\S$ Research Center for Computer Science and Information Technologies, Macedonian
Academy of Sciences and Arts, Bul. Krste Misirkov 2, 1000 Skopje, Macedonia\\
$\pounds$ School of Biological Sciences, Institute for Research in Fundamental
Sciences, POB 19395-5746, Tehran, Iran\\
$\&$ Ikerbasque -- Basque Foundation for Science, Calle de Mar{\'i}a D{\'i}az
de Haro 3, 48013 Bilbao, Basque Country, Spain\\
$\P$ Akhiezer Institute for Theoretical Physics, Kharkov 61108, Ukraine}

\begin{abstract}
The emerging diffusive dynamics in many complex systems shows a
characteristic crossover behaviour from anomalous to normal diffusion which
is otherwise fitted by two independent power-laws. A prominent example for
a subdiffusive-diffusive crossover are viscoelastic systems such as lipid
bilayer membranes, while superdiffusive-diffusive crossovers occur in systems
of actively moving biological cells. We here consider the general dynamics of
a stochastic particle driven by so-called tempered fractional Gaussian noise,
that is noise with Gaussian amplitude and power-law correlations, which are
cut off at some mesoscopic time scale. Concretely we consider such noise with
built-in exponential or power-law tempering, driving an overdamped Langevin
equation (fractional Brownian motion) and fractional Langevin equation motion.
We derive explicit
expressions for the mean squared displacement and correlation functions,
including different shapes of the crossover behaviour depending on the
concrete tempering, and discuss the physical meaning of the tempering. In the
case of power-law tempering we also find a crossover behaviour from faster to
slower superdiffusion and slower to faster subdiffusion. As a direct application
of our model we demonstrate that the obtained dynamics quantitatively described
the subdiffusion-diffusion and subdiffusion-subdiffusion crossover in lipid bilayer
systems. We also show that a model of tempered fractional Brownian motion recently
proposed by Sabzikar and Meerschaert leads to physically very different behaviour
with a seemingly paradoxical ballistic long time scaling.
\end{abstract}

\section{Introduction}

Diffusion, the stochastic motion of a tracer particle, was beautifully
described by Brown in his study of pollen granules and a multitude of
other \emph{molecules\/} (microscopic particles) \cite{brown}. Diffusion is
typically described in terms of the mean squared displacement (MSD)
\begin{equation}
\label{msd}
\langle x^2(t)\rangle\simeq\mathcal{D}_{\alpha}t^{\alpha}
\end{equation}
of the particle spreading. When $\alpha=1$ this is the well known law of
normal (Brownian or Fickian) diffusion observed in detailed quantitative
studies by Perrin, Nordlund, and Kappler \cite{perrin,nordlund,kappler},
among others. In the case of a scaling with an exponent $\alpha$
different from unity,
the dynamics encoded by the MSD (\ref{msd}) can be classified in terms
of the anomalous diffusion exponent $\alpha$ as either subdiffusive for
$0<\alpha<1$ or superdiffusive for $\alpha>1$ \cite{bouchaud,report}. In
expression (\ref{msd}) the generalised diffusion coefficient has physical
dimension $[\mathcal{D}_{\alpha}]=\mathrm{cm}^2/\mathrm{s}^{\alpha}$. Anomalous
diffusion with $\alpha\neq1$ has been revealed in a multitude of systems
\cite{bouchaud,report,pccp}. In particular, following the massive advances
of microscopy techniques anomalous diffusion was discovered in a surging
number of biological systems \cite{hoefling,norregaard}. Thus, subdiffusion
was monitored for both endogenous and introduced submicron tracers in
biological cells \cite{elbaum,seisenhuber,weiss,lene0,garini,lene,
stas,lampo} or in inanimate, artificially crowded systems
\cite{banks,weiss1,lene1}. Supercomputing studies of protein
internal motion \cite{smith} or of constituent molecules of dilute and
protein-crowded lipid bilayer membranes \cite{jae_membrane,ilpo_faraday,
jae_prx,kneller,kneller1} also show subdiffusive behaviour. Due to
active motion, also superdiffusion has been reported from several
cellular systems \cite{elbaum,seisenhuber,granick,christine,jae}. For
a more exhaustive list of systems see the recent reviews
\cite{hoefling,norregaard,ilpo,membrane_rev,yasmine}.

In most of these systems the observed anomalous diffusion was identified
as fractional Brownian motion or fractional Langevin equation motion type
defined below. Both are characterised by power-law correlations of the
driving noise \cite{pccp,hoefling,igor_softer}. At sufficiently long times,
however, this anomalous diffusion will eventually cross over to normal
diffusion, when the system's temporal evolution exceeds some relevant
correlation time. For instance, all atom molecular dynamics simulations of
pure lipid bilayer membranes exhibit a subdiffusive-diffusive crossover
at around 10 nsec, the time scale when two lipids mutually exchange
their position \cite{jae_membrane}. The quantitative description of this
anomalous-to-normal crossover is the topic of this paper. For both the
subdiffusive and superdiffusive situations we include a maximum correlation
time of the driving noise and provide exact solutions for the MSD in the
case of \emph{hard, exponential\/} and \emph{power-law\/} truncation,
so-called \emph{tempering}, that can be easily applied in the analysis
of experimental or simulations data. The advantage of such a model, in
comparison to simply combining an anomalous and a normal diffusive law for
the MSD is that the crossover is built into a two-parameter exponential
tempering model depending only on the noise strength driving the motion and
the crossover time. For the case of a power-law tempering an additional
scaling exponent enters. Depending on its magnitude, the anomalous-normal
crossover dynamics can be extended to a crossover from either faster to
slower superdiffusion or slower to faster subdiffusion. In this approach
the crossover between different diffusion regimes thus naturally emerges,
and the type of tempering governs the exact crossover shape. As we will show
the crossover shape encoded in this approach nicely fits actual data.

The paper is structured as follows. In section \ref{sec_lang} we consider the
tempering of superdiffusive fractional Brownian motion and derive the crossover
to normal diffusion. In section \ref{sec_gle} we perform the same tasks
for the subdiffusive generalised Langevin equation. Section \ref{sec_exp}
compares our subdiffusive to normal diffusive model of the tempered generalised
Langevin equation to supercomputing data from lipid bilayer membranes exhibiting
characteristic crossover dynamics. The data analysis demonstrates excellent
agreement with the built-in crossover behaviour of our model. Section \ref{sec_tfbm}
addresses direct tempering suggested by Meerschaert and Sabzikar as well as its
physicality. Indeed, we show that this type of tempering leads to ballistic
motion. We conclude in section \ref{sec_conc}. Several short appendices provide
some additional mathematical details.

\section{Tempered superdiffusive fractional Brownian motion}
\label{sec_lang}

We start from the overdamped stochastic equation of motion of a physical test
particle in a viscous medium under the influence of a stochastic force $\xi(t)$
\cite{risken,vankampen}
\begin{equation}
\label{LE_overdamped}
\frac{dx(t)}{dt}=\frac{\xi(t)}{m\eta}=v(t),
\end{equation}
where $x(t)$ is the particle position and $v(t)$ its velocity. Without loss of
generality we assume the initial condition $x(0)=0$. Furthermore, $m$
is the particle mass, and $\eta$, of physical dimension $[\eta]=\mathrm{s}^{-1}$
is the friction coefficient. The stochastic force $\xi(t)$ is assumed to be a
stationary and Gaussian noise of zero mean. Then the velocity autocorrelation
function fulfils
\begin{equation}
\label{autocorr}
\langle v(t)v(t+\tau)\rangle=\langle v^2\rangle_{\tau},
\end{equation}
for all $\tau\ge0$. By formal integration of equation (\ref{LE_overdamped}) the
MSD yields in the form
\begin{eqnarray}
\nonumber
\langle x^2(t)\rangle&=&\int_0^tdt_1\int_0^tdt_2\langle v(t_1)v(t_2)\rangle\\
&=&2\int_0^tdt_1\int_{t_1}^tdt_2\langle v(t_1)v(t_2)\rangle=2\int_0^td\tau
(t-\tau)\langle v^2\rangle_{\tau}.
\label{msd_general}
\end{eqnarray}
From this result we infer that if the autocorrelation function $\langle v^2
\rangle_{\tau}$ decays sufficiently fast at long times, such that $\int_0^{\infty}d\tau
\langle v^2\rangle_{\tau}$ is finite, then the MSD reads
\begin{equation}
\label{MSD_normal}
\langle x^2(t)\rangle\sim2t\int_0^{\infty}d\tau\langle v^2\rangle_{\tau},
\end{equation}
at $t\to\infty$, and diffusion becomes asymptotically normal. Thus, one should
expect anomalous diffusion at long times whenever $\int_0^{\infty}d\tau\langle
v^2\rangle_{\tau}$ is either infinity or zero. This is exactly the case for the
persistent and antipersistent fractional Gaussian motions considered in what
follows, respectively. In the case of superdiffusive fractional Brownian motion
we choose the autocorrelation function in the form
\begin{equation}
\label{correlation_power-law}
\langle v^2\rangle_{\tau}=\frac{\mathcal{D}_H}{\Gamma(2H-1)}\tau^{2H-2},
\end{equation}
where the constant noise strength $\mathcal{D}_H$ has dimension $[\mathcal{D}_H]=
\mathrm{cm}^2/\mathrm{s}^{2H}$, $\Gamma(z)$ is the Gamma function, and
the Hurst exponent $H$ is in the interval $1/2\le H<1$. We note here that this
approach leads to the correct power-law asymptotics of the classical
Mandelbrot-van Ness fractional Gaussian noise at long times \cite{mandelbrot}
with $\int_0^{\infty} d\tau\langle v^2\rangle_{\tau}=\infty$, but at the same
time leads to an infinite zero-point variance $\langle v^2\rangle_{\tau=0}$ of the
noise.\footnote{A more consistent approach using the smoothening procedure of
fractional Brownian motion over infinitesimally small time intervals {\`a} la
Mandelbrot and van Ness
\cite{mandelbrot} shows that the weak divergence of the autocorrelation function
(\ref{correlation_power-law}) at $\tau=0$ does not lead to a change of the MSD.}
Keeping away from $\tau=0$ we are allowed to restrict ourselves to the power-law form
(\ref{correlation_power-law}). Furthermore the coefficient $\Gamma(2H-1)$ in
equation (\ref{correlation_power-law}) is introduced to capture the white noise
limit. Indeed, due to the property of the $\delta$-function \cite{gelfand and shilov}
\begin{equation}
\label{delta limit}
\lim_{H\to0.5^+}\frac{\tau^{2H-2}}{\Gamma(2H-1)}=\delta(\tau)
\end{equation}
at $H=0.5$ and with $\int_0^{\infty}d\tau\delta(\tau)=1$ equation
(\ref{correlation_power-law}) reduces to
\begin{equation}
\label{delta_correlation}
\langle v^2\rangle_{\tau}=\mathcal{D}\delta(\tau)
\end{equation}
with $\mathcal{D}_{1/2}=\mathcal{D}$.\footnote{The power-law correlations in
the autocorrelation function (\ref{correlation_power-law}) contrast the sharp
$\delta$-correlation of relation (\ref{delta_correlation}) \cite{zwanzig,coffey}.
We note that in this combination of the Langevin equation (\ref{LE_overdamped})
and the autocorrelation function (\ref{correlation_power-law}) the fluctuation
dissipation theorem is not satisfied, and the noise $\xi(t)$ can be considered
as an external noise \cite{klimontovich}, see also the discussion of the
generalised Langevin equation below.}

Now, after plugging result (\ref{correlation_power-law}) into expression
(\ref{msd_general}) the MSD can be readily calculated, yielding
\begin{equation}
\label{msd_fractional_noise}
\langle x^{2}(t)\rangle=\frac{2\mathcal{D}_H}{\Gamma(2H+1)}t^{2H},
\end{equation}
which yields sub-ballistic superdiffusion with the anomalous diffusion exponent
$\alpha=2H$, and thus $1<\alpha<2$.

In what follows we consider both a hard exponential and a power-law
truncation (tempering) of the persistent fractional Gaussian noise with Hurst
exponent $1/2\le H<1$.

\subsection{Exponentially truncated fractional Gaussian noise}

Let us first consider an exponential tempering of the form
\begin{equation}
\label{correlation_power-law_exp_truncation}
\langle v^2\rangle_{\tau}=\frac{\mathcal{D}_H}{\Gamma(2H-1)}\tau^{2H-2}e^{-\tau/
\tau_{\star}},
\end{equation}
for $\tau>0$, where $\tau_{\star}>0$ is a characteristic crossover time scale.
For instance, in the case of moving cells the crossover time $\tau_{\star}$ would
correspond to the time scale when the cell motion becomes uncorrelated, similar to
the decorrelation of the lipid motion in the example of the lipid bilayer system
discussed below.

Here we note that one should keep in mind that the autocorrelation function
$\langle v^2\rangle_{\tau}$ can not be chosen arbitrary. Namely, its Fourier
transform, the spectrum $\langle\tilde{v}^2(\omega)\rangle$ of the random
process $v(t)$ must be non-negative \cite{specref}. The positivity of $\langle
\tilde{v}^2(\omega)\rangle$ for the case of exponential tempering in equation
(\ref{correlation_power-law_exp_truncation}) is shown in \ref{appa}.
Note also that now $\int_0^{\infty}d\tau\langle v^2\rangle_{\tau}=\mathcal{D}_H
\tau_{\star}^{2H-1}$ is finite, thus we expect normal diffusion at long times.

With the use of expression (\ref{msd_general}) the MSD for the exponentially
truncated fractional Gaussian noise takes on the exact form
\begin{equation}
\label{msd_fgn_exp}
\langle x^{2}(t)\rangle=\frac{2\mathcal{D}_H\tau_{\star}^{2H}}{\Gamma(2H-1)}
\left[\frac{t}{\tau_{\star}}\gamma\left(2H-1,\frac{t}{\tau_{\star}}\right)-
\gamma\left(2H,\frac{t}{\tau_{\star}}\right)\right],
\end{equation}
where $\gamma(a,z)=\int_0^zt^{a-1}e^{-t}dt$ is the incomplete $\gamma$-function.
Using the asymptotic $\gamma(a,z)\sim z^{a}/a$ for $z\ll1$, and $\gamma(a,z)\sim
\Gamma(a)$ for $z\gg1$, we observe superdiffusive behaviour at short times, and
normal diffusion at long times, namely,
\begin{equation}
\label{msd_fgn_exp_long}
\langle x^2(t)\rangle\sim\left\{\begin{array}{ll}
\frac{\displaystyle2\mathcal{D}_H}{\displaystyle\Gamma(2H+1)}t^{2H},&t\ll\tau_{
\star}\\[0.6cm]
2\mathcal{D}_H\tau_{\star}^{2H-1}t,&t\gg\tau_{\star}.\end{array}\right.
\end{equation}
The emerging normal diffusion thus has the effective diffusivity $\mathcal{D}_H\tau_{
\star}^{2H-1}$. Note that the approximate formula at long times is in concordance
with the simple estimate given by expression (\ref{MSD_normal}).

\begin{figure}
\centering
\includegraphics[width=10cm]{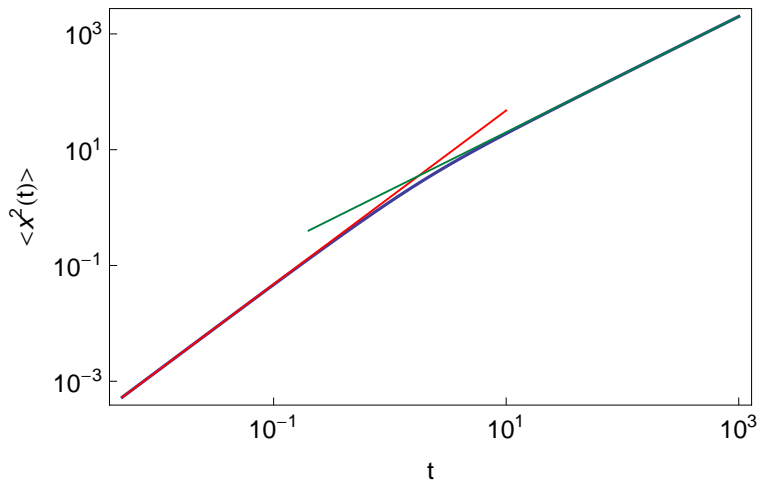}
\caption{Mean squared displacement (\ref{msd_fgn_exp}) for superdiffusive
fractional Brownian motion with $H=3/4$, $\mathcal{D}_H=1$,
and $\tau_{\star}=1$ (blue line). The short and long time asymptotics
given by expression (\ref{msd_fgn_exp_long}) are depicted by the red and
green lines, respectively.}
\label{fgr:LEexptrunc}
\end{figure}

Figure \ref{fgr:LEexptrunc} shows the crossover behaviour from superdiffusion
to normal diffusion encoded in expression (\ref{msd_fgn_exp}), along with the
short and long time asymptotes given by result (\ref{msd_fgn_exp_long}). As can
be discerned from the plot, the crossover region is fairly short, spanning less
than a decade in time for the chosen parameters.

\subsection{Power-law truncated fractional Gaussian noise}

We now consider the softer power-law truncation of the form
\begin{equation}
\label{correlation_power-law_power-law_truncation}
\langle v^2\rangle_{\tau}=\frac{\mathcal{D}_H}{\Gamma(2H-1)}\tau^{2H-2}\left(
1+\frac{\tau}{\tau_{\star}}\right)^{-\mu},
\end{equation}
for $\tau>0$, $\mu>0$ and compare the resulting behaviour with the scenario of
exponential tempering. Here, apart from the crossover time $\tau_{\star}$ the
new power-law exponent $\mu$ is introduced which effects the dynamics at long
times, as we are going to show below. We remark that the positivity of the
spectrum for the power-law truncated form is discussed in \ref{appa}.
After plugging (\ref{correlation_power-law_power-law_truncation})
into expression (\ref{msd_general}) we find for the MSD that
\begin{equation}
\label{msd_step}
\langle x^{2}(t)\rangle=\frac{2\mathcal{D}_H\tau_{\star}^{2H}}{\Gamma(2H-1)}
\left[\frac{t}{\tau_{\star}}f\left(\mu,2H-1;\frac{t}{\tau_{\star}}\right)-f\left(
\mu,2H;\frac{t}{\tau_{\star}}\right)\right],
\end{equation}
where we introduced the notation
\begin{equation}
\label{aux}
f(\mu,\alpha;a)=\int_0^a\frac{y^{\alpha-1}}{(1+y)^{\mu}}dy.
\end{equation}
Now, using the integral representation \cite{Abramowitz} of
the hypergeometric function $_2F_1$ \cite{erdelyi} we rewrite the integral in
equation (\ref{aux}) as
\begin{equation}
f(\mu,\alpha;a)=\frac{a^{\alpha}}{\alpha}{_2}F_1(\mu,\alpha,\alpha+1;-a),
\end{equation}
and thus rewrite the MSD (\ref{msd_step}) in the final form
\begin{eqnarray}
\nonumber
\langle x^{2}(t)\rangle&=&\frac{2\mathcal{D}_Ht^{2H}}{\Gamma(2H-1)}\left[
\frac{1}{2H-1}{_{2}}F_{1}\left(\mu,2H-1;2H;-\frac{t}{\tau_{\star}}\right)\right.\\
&&\hspace*{2.2cm}\left.-\frac{1}{2H}{_{2}}F_{1}\left(\mu,2H;2H+1;-\frac{t}{\tau_{
\star}}\right)\right],
\label{msd fractional noise power-law truncation}
\end{eqnarray}
In this notation the MSD can be directly evaluated by Wolfram Mathematica
\cite{mathematica}. Note
that $_{2}F_{1}(0,b;c;z)=1$, and thus result (\ref{msd fractional noise power-law
truncation}) reduces exactly to the MSD (\ref{msd_fractional_noise}) for the
untruncated case $\mu=0$. To obtain the limiting behaviours of the MSD
(\ref{msd fractional noise power-law truncation}) at short times $t\ll\tau_{\star}$
we use the Gauss hypergeometric series for the function ${}_{2}F_{1}$, see 15.1.1
in \cite{Abramowitz}. As result, to leading order we recover the MSD
(\ref{msd_fractional_noise}) of untruncated fractional Brownian motion.

At long times $t\gg\tau_{\star}$ the situation for power-law tempering is actually
richer than for the case of exponential tempering. To see this, we first employ
the linear transformation formula 15.3.7 in \cite{Abramowitz} and write expression
(\ref{msd fractional noise power-law truncation}) in the form
\begin{eqnarray}
\nonumber
\fl\langle x^2(t)\rangle&=&\frac{2D_H\tau_{\star}^{2H-1}t}{\Gamma(2H-1)}\left[\frac{
\Gamma(2H-1)\Gamma(\mu+1-2H)}{\Gamma(\mu)}-\frac{\Gamma(2H+1)\Gamma(\mu-2H)}{2H\Gamma
(\mu)}\frac{\tau_{\star}}{t}\right.\\
\nonumber
\fl&&+\frac{1}{2H-\mu-1}\left(\frac{\tau_{\star}}{t}\right)^{\mu+1-2H}\,
_2F_1\left(\mu,\mu+1-2H;\mu+2-2H;-\frac{\tau_{\star}}{t}\right)\\
\fl&&\left.-\frac{1}{2H-\mu}\left(\frac{\tau_{\star}}{t}\right)^{\mu+1-2H}\,
_2F_1\left(\mu,\mu-2H;\mu+1-2H;-\frac{\tau_{\star}}{t}\right)\right].
\label{fexp}
\end{eqnarray}
We consider two possible cases:

\subsubsection{Weak power-law truncation, $0<\mu<2H-1<1$.}

In this case the third and fourth terms in the square brackets of expression
(\ref{fexp}) are dominating and we find
\begin{equation}
\label{msdweak}
\langle x^2(t)\rangle\sim\frac{2D_H\tau_{\star}^{\mu}}{(2H-\mu)(2H-1-\mu)\Gamma(
2H-1)}t^{2H-\mu}
\end{equation}
for $t\gg\tau_{\star}$. Note that in the limit $\mu\to0$ result (\ref{msdweak})
reduces to the untruncated formula (\ref{msd_fractional_noise}). Thus, since
we observe the inequality $2H-\mu>1$ in the case of weak power-law truncation
the dynamics is still superdiffusive, however, with a reduced anomalous diffusion
exponent smaller than the value $2H$ in the short time limit.

\subsubsection{Strong power-law truncation, $\mu>2H-1>0$.}

Note that in this case the integral of the velocity autocorrelation function
(\ref{correlation_power-law_power-law_truncation}) over the whole time domain
converges,  $\int_0^{\infty}d\tau\langle v^2\rangle_{\tau}=\mathcal{D}_H\tau_{
\star}^{2H-1}\Gamma(\mu-2H+1)/\Gamma(\mu)$, see 2.2.5.24 in \cite{prudnikov1}.
Thus, with expression (\ref{MSD_normal}) we expect a linear time behaviour in
the long time limit, whereas the term to next order in (\ref{msd_general}) gives
$\int^td\tau\tau\langle v^2\rangle_{\tau}\simeq\int^td\tau\tau^{2H-1-\mu}\simeq
t^{2H-\mu}$, a sublinear contribution since $2H-\mu<1$. Alternatively, it follows
from (\ref{fexp}) that the main contribution comes from the first term in the
square brackets. Thus, in full accordance with expression (\ref{MSD_normal}) we
get
\begin{equation}
\label{msd fractional noise power-law truncation long}
\langle x^2(t)\rangle\sim
\frac{\displaystyle2\mathcal{D}_H\Gamma(\mu-2H+1)\tau_{\star}^{2H-1}t}{
\displaystyle\Gamma(\mu)}
\end{equation}
at $t\gg\tau_{\star}$.

Finally, for the borderline case $0<\mu=2H-1<1$ it is in fact easier to consider
equation (\ref{msd fractional noise power-law truncation}). Making use of formula
7.3.1.81 in \cite{prudnikov2} we see that the leading contribution comes from the
first hypergeometric function in the square brackets in expression 
(\ref{msd fractional noise power-law truncation}), as $_2F_1(2H-1,2H-1;2H;z)\sim
\Gamma(2H)\Gamma^{-1}(2H-1)(-z)^{-2H+1}\ln(-z)$. For the MSD we then finally
obtain
\begin{equation}
\langle x^2(t)\rangle\sim\frac{2D_H\tau_{\star}^{2H-1}}{\Gamma(2H-1)}t\ln\left(
\frac{t}{\tau_{\star}}\right).
\end{equation}
Thus, in this borderline limit between weak truncation (leading to reduced
superdiffusion at long times) and strong truncation (normal long time diffusion)
we here obtain normal diffusion with a logarithmic correction.

\begin{figure}
\centering
\includegraphics[width=10cm]{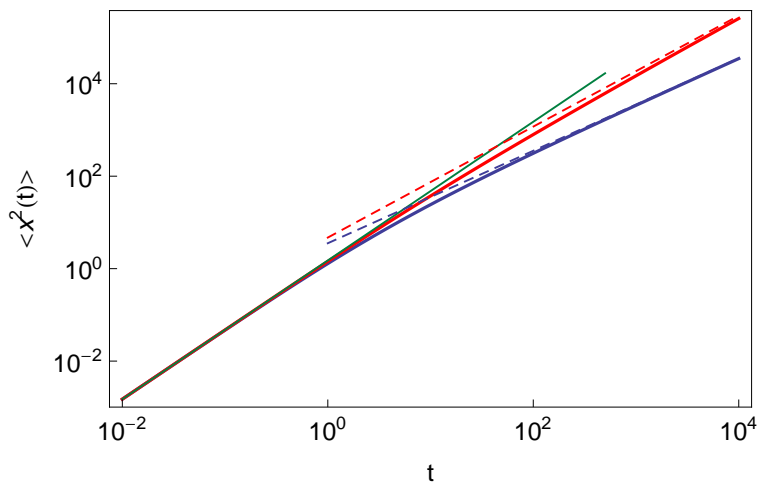}
\caption{MSD (\ref{msd fractional noise power-law truncation}) for power-law
tempered fractional Brownian motion with $H=3/4$, $\mathcal{D}
_H=1$, and $\tau_{\star}=1$. The red solid line is for $\mu=0.3$ (weak power
law truncation), whereas the blue solid line is for $\mu=1$ (strong power-law
truncation). The red and blue dashed lines correspond to the asymptotics
(\ref{msdweak}) and (\ref{msd fractional noise power-law truncation long}),
respectively. The behaviour for the untruncated case given by expression
(\ref{msd_fractional_noise}) is depicted by the green solid line.}
\label{fgr:LEpowertrunc}
\end{figure}

Figure \ref{fgr:LEpowertrunc} demonstrates that for the power-law tempering the
crossover region is significantly enhanced, spanning several orders of magnitude,
as compared to the much swifter crossover in the case of exponential tempering.

\begin{figure}
\centering
\includegraphics[width=10cm]{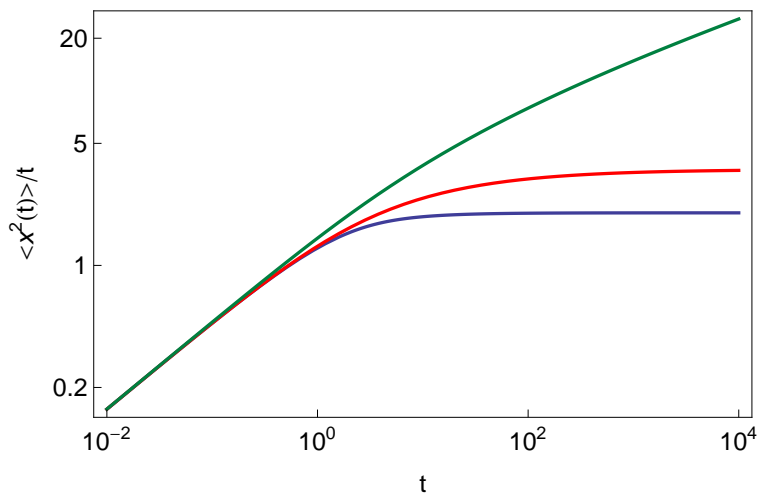}
\caption{Comparison of the ratio $\langle x^2\rangle/t$ for different modes of
truncation of the power-law noise in equation (2). Parameters: $H=3/4$,
$D_H=1$, and $\tau_{\star}=1$. From bottom to top the blue line depicts the
exponential truncation (\ref{msd_fgn_exp}) while the red line and green lines
show expression (\ref{msd fractional noise power-law truncation}) for strong
($\mu=1$) and weak ($\mu=0.3$) power-law truncation, respectively.}
\label{fgr:LEexpVSpower}
\end{figure}

The MSDs for both cases of exponential and power-law truncation are directly
compared in figure \ref{fgr:LEexpVSpower}, along with the time derivative of
the MSD. As can be seen, the crossover for the exponential tempering occurs
much more rapidly. Thus also the amplitude of the long time Brownian scaling
is higher in the case of the power-law tempering for the same value of the
crossover time scale $\tau_{\star}$.

\begin{figure}
\centering
\includegraphics[width=10cm]{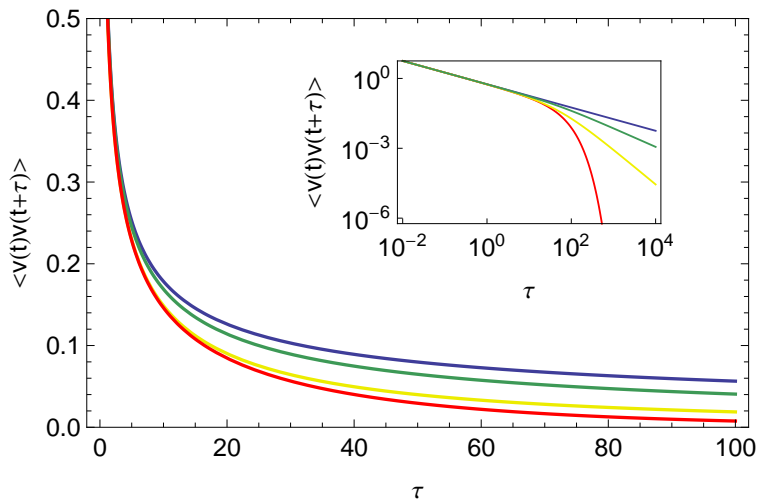}
\caption{Main figure. Comparison of the velocity autocorrelation functions,
from top to bottom: untruncated motion, equation (\ref{correlation_power-law})
(blue line), weak power-law truncation, equation
(\ref{correlation_power-law_power-law_truncation}) with $\mu=0.3$ (green line),
strong power-law truncation, equation
(\ref{correlation_power-law_power-law_truncation}) with $\mu=1$ (yellow line),
and exponential truncation, equation (\ref{correlation_power-law_exp_truncation})
(red line). Parameters: $H=3/4$, $D_H=1$, and $\tau_{\star}=50$.
Inset: double-logarithmic representation.}
\label{fgr:fig_corr}
\end{figure}

A graphical representation of the correlation functions (\ref{correlation_power-law}),
(\ref{correlation_power-law_exp_truncation}) and
(\ref{correlation_power-law_power-law_truncation}) is given in figure
\ref{fgr:fig_corr}. The exponential cutoff appears more abrupt, as it should.
However, this difference will obviously be reduced for larger values of the cutoff
exponent $\mu$. To fit data, the crossover shape can thus be adjusted by the choice
of $\mu$ for the case of power-law tempering, thus having the possibility to effect
a gradual adjustment from soft power-law to hard exponential tempering.

\section{Tempered subdiffusive generalised Langevin equation motion}
\label{sec_gle}

We now consider the motion encoded in the overdamped generalised Langevin equation
for a particle with mass $m$ moving in a viscous medium characterised by the
friction kernel $\gamma_H(t)$ of dimension $[\gamma_H(t)]=\mathrm{s}^{-2}$
\cite{pccp,zwanzig,kubo}
\begin{equation}
\label{GLE_overdamped}
m\int_0^t\gamma_H(t-t')\frac{dx(t')}{dt'}dt'=\xi(t),
\end{equation}
where $x(0)=0$ without loss of generality. Similar to the model considered in
section \ref{sec_lang} $\xi(t)$ is a Gaussian noise with power-law correlation
of the form (\ref{correlation_power-law}) with $1/2\le H<1$. However, in contrast
to the fractional Brownian motion model considered above, we require the system
to be thermalised, such that the random force is coupled to the friction kernel
through the Kubo-Zwanzig fluctuations dissipation relation \cite{zwanzig,kubo}
\begin{equation}
\label{correlation1}
\langle\xi^2\rangle_{\tau}=k_BTm\gamma_H(\tau).
\end{equation}

\subsection{Mean squared displacement}

Let us recall the derivation of the MSD from equations (\ref{GLE_overdamped}) and
(\ref{correlation1}).
With our choice $x(0)=0$ we obtain for the Laplace transform of $x(t)$,
$\tilde{x}(s)=\int_0^{\infty}x(t)\exp(-st)dt$ that
\begin{equation}
\tilde{x}(s)=\frac{\tilde{\xi}(s)}{ms\tilde{\gamma}_H(s)}.
\end{equation}
Inverse Laplace transformation produces
\begin{equation}
x(t)=\frac{1}{m}\int_0^t\xi(t')H(t-t')dt',
\end{equation}
where the kernel $H(t)$ is the inverse Laplace transform of $\tilde{H}(s)=1/[
s\tilde{\gamma}_H(s)]$. After some transformation we recover the MSD
\begin{eqnarray}
\nonumber
\langle x^2(t)\rangle&=&\frac{2}{m^2}\int_0^tdt_1\int_{t_1}^tdt_2H(t-t_1)H(t-t_2)
\langle\xi^2\rangle_{t_2-t_1}\\
&=&\frac{2k_BT}{m}\int_0^tH(t')M(t')dt',
\end{eqnarray}
where we introduced $M(t)=\int_0^t\gamma_H(t')H(t-t')dt'$. Its Laplace
transform is $\tilde{M}(s)=\tilde{\gamma}_H(s)\tilde{H}(s)=1/s$, and thus
simply $M(t)=1$. We therefore arrive at
\begin{equation}
\langle x^2(t)\rangle=\frac{2k_BT}{m}\int_0^tH(t')dt'.
\end{equation}
In Laplace space, this relation reads
\begin{equation}
\label{lapmsd}
\langle\tilde{x^2}(s)\rangle=\frac{2k_BT}{m}\frac{\tilde{H}(s)}{s}=\frac{2k_BT}{m}
\frac{1}{s^2\tilde{\gamma}_H(s)}.
\end{equation}

We stop to include a note on when exactly we expect asymptotically normal
diffusion in the generalised Langevin equation model. The reasoning is similar
to that presented at the beginning of section \ref{sec_lang}. Namely, from
equation (\ref{lapmsd}) it follows that diffusion is normal at long times if $\tilde{
\gamma}_H(s)$ tends to a constant in the limit $s\to0$. This is equivalent
to requiring that the average $\int_0^{\infty}\gamma_H(\tau)d\tau$ is finite
or, taking into account the fluctuation-dissipation relation (\ref{correlation1})
that $\int_0^{\infty}\langle\xi^2\rangle_{\tau}d\tau$ is finite (similar to
the conclusion in section \ref{sec_lang}). Then, from expression (\ref{lapmsd})
we infer the following behaviour in the long time limit (compare with equation
(\ref{MSD_normal}))
\begin{equation}
\label{msdlong}
\langle x^2(t)\rangle=\frac{2k_BT}{m\int_0^{\infty}\gamma_H(\tau)d\tau}t.
\end{equation}
According to this, anomalous diffusion is expected at long times whenever
$\int_0^{\infty}\gamma_H(\tau)d\tau$ is either infinite (subdiffusion) or
zero (superdiffusion).\footnote{Note here the difference to the results in
section \ref{sec_lang} where the fluctuation-dissipation theorem is not applied:
in that case divergence of the integral over the correlator of the noise $\xi(t)$
over the entire time domain leads to superdiffusion, while subdiffusion emerges
when the integral is identical to zero.}

In accordance with section \ref{sec_lang} we choose the friction kernel in the
power-law form
\begin{equation}
\label{plcorr}
\gamma_H(\tau)=\frac{\Gamma_H}{\Gamma(2H-1)}\tau^{2H-2},
\end{equation}
where the coefficient $\Gamma_H$ is of dimension $[\Gamma_H]=\mathrm{s}^{-2H}$.
The normal Brownian case is recovered from equation (\ref{GLE_overdamped}) for
$H=1/2$ since for $H\to1/2+$ we see that $\gamma_H(t)\to\Gamma_{1/2}\delta(t)$
(note that in this Brownian limit, $\Gamma_{1/2}=\eta$)
and equation (\ref{GLE_overdamped}) assumes the form of the standard Langevin
equation driven by white Gaussian noise obeying the regular fluctuation
dissipation theorem. We note that the memory kernel for the power-law form
(\ref{plcorr}) can be rewritten in terms of a fractional derivative, and the
resulting version of equation (\ref{GLE_overdamped}) is then often referred to
as the fractional Langevin equation \cite{pccp,deng,lutz,goychuk}. Power-law
memory kernels of the form (\ref{plcorr}) are typical for many viscoelastic
systems \cite{hoefling,norregaard,garini,lene,stas,lampo,weiss1,lene1,
jae_membrane,goychuk}.

We now use the Laplace transform of equation (\ref{plcorr}), $\tilde{\gamma}_H(s)=
\Gamma_Hs^{1-2H}$, plug this into the above expression, and take an inverse
Laplace transformation. This procedure leads to the final result
\begin{equation}
\label{msd fractional noise gle}
\langle x^2(t)\rangle=\frac{1}{\Gamma(3-2H)}\frac{2k_BT}{m\Gamma_H}t^{2-2H},
\end{equation}
which reduces to the classical result $\langle x^2(t)\rangle=2(k_BT/[m\eta])t$
for normal Brownian motion in the limit $H=1/2$. Therefore, due to the requirement
that the system is thermalised and thus the Kubo-Zwanzig fluctuation theorem is
fulfilled, the same noise leads to subdiffusion in this case with anomalous
diffusion exponent $\alpha=2-2H$ and $0<2-2H<1$. Indeed, due to the coupling in
relation (\ref{correlation1}) large noise values lead to large friction values,
and therefore the persistence of the noise is turned into antipersistent diffusion
dynamics \cite{pccp,deng,goychuk}.

\subsection{Autocorrelation functions of displacements and velocities}

We now derive the autocorrelation function of the displacements, following the
procedure laid out by Pottier \cite{pottier}. First, we note that the double Laplace
transform of the correlation function of the random force can be written as
\begin{equation}
\langle\tilde{\xi}(s_1)\tilde{\xi}(s_2)\rangle=k_BTm\int_0^{\infty}dt_1\int_0^
{\infty}dt_2e^{-s_1t_1-s_2t_2}\gamma_H(|t_2-t_1|).
\end{equation}
Then we split the domain of integration over $t_2$ into the two domains $0\le
t_2\le t_1$ and $t_1\le t_2<\infty$. After introducing $\tau=t_1-t_2$ and $\tau
=t_2-t_1$ in each domain, respectively, we arrive at
\begin{equation}
\label{aux1}
\langle\tilde{\xi}(s_1)\tilde{\xi}(s_2)\rangle=k_BTm\frac{\tilde{\gamma}_H(s_1)
+\tilde{\gamma}_H(s_2)}{s_1+s_2}.
\end{equation}
This expression represents the Laplace domain formulation of the fluctuation
dissipation theorem (\ref{correlation1}). By help of equations (\ref{aux1}) and
(\ref{GLE_overdamped}) we then obtain the double Laplace transform of the
displacement correlation function,
\begin{equation}
\label{auxcorr}
\langle\tilde{x}(s_1)\tilde{x}(s_2)\rangle=\frac{k_BT}{m}\left(\frac{1/\tilde{
\gamma}_H(s_1)}{s_1s_2(s_1+s_2)}+\frac{1/\tilde{\gamma}_H(s_2)}{s_1s_2(s_1+s_2)}
\right).
\end{equation}
In the first term in the parentheses we first take the inverse Laplace
transformation over $s_2$, going from $1/[s_2(s_1+s_2)]$ to $[1-\exp(-s_1t_2)]/
s_1$. Exchanging $s_2$ for $s_1$ we perform the same operation on the second
term. Then we inverse Laplace transform the first term with respect to $s_1$
and make use of the translation formula $\mathscr{L}_s^{-1}\left\{\exp(-bs)
\mathscr{L}_s\left\{f(t)\right\}\right\}=f(t-b)\Theta(t-b)$, where $b>0$ and
$\Theta(t)$ is the Heaviside step function. As result yields
\begin{equation}
\langle x(t_1)x(t_2)\rangle=\frac{1}{\Gamma(3-2H)}\frac{k_BT}{m\Gamma_H}
\Big(t_1^{2-2H}+t_2^{2-2H}-|t_2-t_1|^{2-2H}\Big).
\end{equation}
The velocity autocorrelation function is obtained by differentiation of this
expression,
\begin{equation}
\label{vcorr}
\langle v(t_1)v(t_2)\rangle=\langle v^2\rangle_{\tau}=-\frac{\sin(\pi[2H-1])
\Gamma(2H)}{\pi}\frac{k_BT}{m\Gamma_H}|\tau|^{-2H},
\end{equation}
where $\tau=t_2-t_1$. We see that in the relevant parameter range $1/2<H<1$ the
velocity autocorrelation is negative, $\langle v^2\rangle_{\tau}<0$, reflecting
the antipersistent character of the resulting motion.

\subsection{Exponentially truncated fractional Gaussian noise}

For the exponentially truncated friction kernel and thus noise autocorrelation
\begin{equation}
\label{gamexp}
\gamma_H(\tau)=\frac{\langle\xi^2\rangle_{\tau}}{k_BTm}=\frac{\Gamma_H}{\Gamma(2H-1)}
\tau^{2H-2}e^{-\tau/\tau_{\star}}
\end{equation}
we obtain the corresponding Laplace transform
\begin{equation}
\label{gamlap}
\tilde{\gamma}_H(s)=\Gamma_H\left(s+\tau_{\star}^{-1}\right)^{1-2H}.
\end{equation}
After plugging this expression into relation (\ref{lapmsd}) and taking the
inverse Laplace transformation we obtain
\begin{equation}
\label{msd exp truncation gle}
\langle x^2(t)\rangle=\frac{2k_BT}{m\Gamma_H}t^{2-2H}E_{1,3-2H}^{1-2H}\left(-
\frac{t}{\tau_{\star}}\right)
\end{equation}
in terms of the three parameter Mittag-Leffler function $E_{\alpha,\beta}^{
\delta}(z)$ (see \ref{appb} for its definition and some relevant properties).
When the crossover time $\tau_{\star}$ tends to infinity, $E_{\alpha,\beta}^
{\delta}(0)=1/\Gamma(\beta)$, and we arrive at result (\ref{msd fractional noise
gle}) for the untruncated noise. In the limit $H=1/2$ we have $\delta=0$ and
$E_{1,2}^0(z)=1/\Gamma(2)=1$, such that equation (\ref{msd exp truncation gle})
reduces to the MSD of normal Brownian motion.

At short times $t\ll\tau_{\star}$ the MSD (\ref{msd exp truncation gle}) reduces
to the subdiffusive expression (\ref{msd fractional noise gle}), whereas at
long times $t\gg\tau_{\star}$ with the help of $E_{1,3-2H}^{1-2H}(-t/\tau_{\star})
\sim(t/\tau_{\star})^{2H-1}$ (see \ref{appa}), in accordance with relation
(\ref{msdlong}) the MSD exhibits normal Brownian behaviour,
\begin{equation}
\label{msd exp truncation gle long}
\langle x^2(t)\rangle\sim\frac{2k_BT}{m\Gamma_H\tau_{\star}^{2H-1}}t.
\end{equation}
We note that a similar crossover was observed in \cite{trifce} where a modified
three-parameter Mittag-Leffler form for the kernel $\gamma_H(\tau)$ was considered.

\begin{figure}
\centering
\includegraphics[width=10cm]{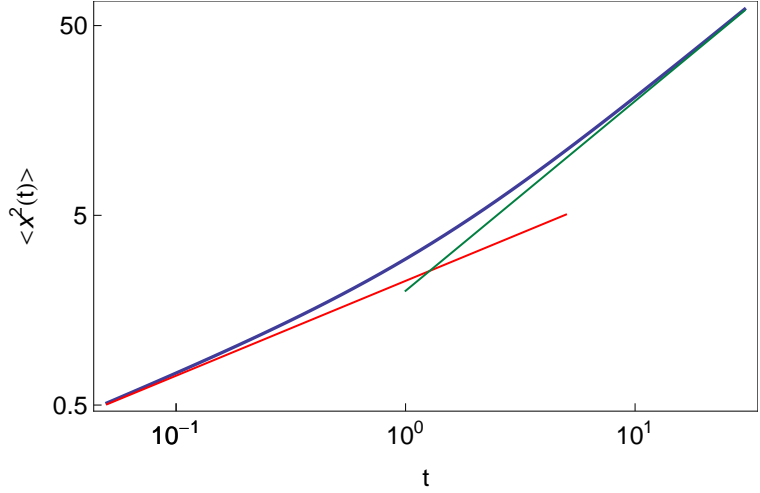}
\caption{MSD (\ref{msd exp truncation gle}) for exponentially tempered generalised
Langevin equation motion with $H=3/4$, $k_BT/[m\Gamma_H]=1$, and
$\tau_{\star}=1$ (blue line). The short and long time asymptotics
(\ref{msd fractional noise gle}) and (\ref{msd exp truncation gle long}) are shown
by the red and green lines, respectively.}
\label{fgr:GLEexptrunc}
\end{figure}

The crossover from subdiffusion to normal diffusion in this exponentially tempered
generalised Langevin equation picture is shown in figure \ref{fgr:GLEexptrunc}. The
crossover behaviour occurs over an interval of the order of a decade in time for
the chosen parameters.

Let us now turn to the autocorrelation functions. Using expression (\ref{gamlap})
in equation (\ref{auxcorr}) we obtain
\begin{eqnarray}
\nonumber
\langle\tilde{x}(s_1)\tilde{x}(s_2)\rangle=\frac{k_BT}{m\Gamma_H}&\left(\frac{1}{
s_1s_2(s_1+s_2)\left(s_1+\tau_{\star}^{-1}\right)^{1-2H}}\right.\\
&\left.+\frac{1}{s_1s_2(s_1+s_2)\left(s_2+\tau_{\star}^{-1}\right)^{1-2H}}\right).
\end{eqnarray}
As above, in the first term in the parentheses we take an inverse Laplace
transformation with respect to $s_2$, and over $s_1$ in the second term.
Then, with the translation formula and the Laplace transform (\ref{laplaceML})
of the three parameter Mittag-Leffler function, we find
\begin{eqnarray}
\nonumber
\langle x(t_1)x(t_2)\rangle=\frac{k_BT}{m\Gamma_H}&\left(t_1^{2-2H}E_{1,3-2H}^{
1-2H}\left(-\frac{t_1}{\tau_{\star}}\right)+t_2^{2-2H}E_{1,3-2H}^{1-2H}\left(-
\frac{t_2}{\tau_{\star}}\right)\right.\\
&\left.-|t_2-t_1|^{2-2H}E_{1,3-2H}^{1-2H}\left(-\frac{|t_2-t_1|}{\tau_{\star}}
\right)\right).
\end{eqnarray}
Differentiation over $t_1$ and $t_2$ (with the help of equation (\ref{mldiff}))
then produces the velocity autocorrelation function,
\begin{equation}
\langle v(t_1)v(t_2)\rangle=\langle v^2\rangle_{\tau}=\frac{k_BT}{m\Gamma_H
\tau^{2H}}E_{1,1-2H}^{1-2H}\left(-\frac{\tau}{\tau_{\star}}\right).
\end{equation}
with $\tau=t_2-t_1>0$. Using the definition (\ref{ML_three}) of the three
parameter Mittag-Leffler function it is easy to check that $E_{1,\delta}^{
\delta}(z)=\exp(z)/\Gamma(\delta)$. Thus, for the velocity autocorrelation
function we find the result
\begin{equation}
\label{vacf exp trunc}
\langle v^2\rangle_{\tau}=-\frac{\sin(\pi[2H-1])\Gamma(2H)}{\pi}\frac{k_BT}{
m\Gamma_H}\tau^{-2H}e^{-\tau/\tau_{\star}},
\end{equation}
which is anticorrelated and reduces to the untruncated result (\ref{vcorr})
when the crossover time $\tau_{\star}$ tends to infinity.

\subsection{Power-law truncated fractional noise}

For the power-law truncated friction kernel and noise autocorrelator,
\begin{equation}
\label{powtrunc}
\gamma_H(\tau)=\frac{\langle\xi^2\rangle_{\tau}}{k_BTm}=\frac{\Gamma_H}{\Gamma(2H-1)}
\tau^{2H-2}\left(1+\frac{\tau}{\tau_{\star}}\right)^{-\mu}
\end{equation}
with $\tau>0$, $\mu>0$ the Laplace transform of the memory kernel can be performed
by use of the integral representation of the Tricomi hypergeometric function
$U(a,b;z)$ (see 13.2.5 of \cite{Abramowitz}), leading to
\begin{equation}
\label{tricomi}
\tilde{\gamma}_H(s)=\Gamma_H\tau_{\star}^{2H-1}U(2H-1,2H-\mu;s\tau_{\star}).
\end{equation}
With the general relation (\ref{lapmsd}) we thus have
\begin{equation}
\label{msd power-law truncation gle}
\langle x^2(t)\rangle=\frac{2k_BT}{m\Gamma_H\tau_{\star}^{2H-1}}g(t)
\end{equation}
with the abbreviation
\begin{equation}
\label{abbrevg}
g(t)=\mathscr{L}^{-1}_s\left\{\frac{1}{s^2U(2H-1,2H-\mu;s\tau_{\star})}\right\}.
\end{equation}
The inverse Laplace transform of expression (\ref{msd power-law truncation gle})
cannot be performed analytically. However,
we make use of the Tauberian theorems\footnote{The Tauberian theorems state that
for slowly varying function $L(t)$ at infinity, i.e. $\lim_{t\rightarrow\infty}
\frac{L(at)}{L(t)}=1$, $a>0$, if $\hat{r}(s)\simeq s^{-\rho}L\left(\frac{1}{s}
\right)$, for $s\rightarrow0$, $\rho\geq0$, then $r(t)=\mathcal{L}^{-1}\left[
\hat{r}(s)\right](t)\simeq\frac{1}{\Gamma(\rho)}t^{\rho-1}L(t),\quad
t\rightarrow\infty$. A similar statement holds for $t\to0$.}
to find the MSD at short and long times.

At short times with $s\tau_{\star}\gg1$ we use the large argument asymptotic of
the Tricomi function, $U(2H-1,2H-\mu;s\tau_{\star})\sim(s\tau_{\star})^{1-2H}$
(13.5.2 in \cite{Abramowitz}) and thus $\tilde{\gamma}_H(s)\sim\Gamma_Hs^{1-2H}$.
From equation (\ref{lapmsd}) (or, equivalently, equations
(\ref{msd power-law truncation gle}) and (\ref{abbrevg}))
we then get to result (\ref{msd fractional noise gle})
by use of the Tauberian theorem.

Similar to the case considered in section \ref{sec_lang}
at long times corresponding to $s\tau_{\star}\ll1$ the situation is actually
richer than for the case of exponential tempering. To see this we first make
use of (13.1.3) in \cite{Abramowitz} to express the Tricomi function via the
Kummer function $M(a,b;z)$ through
\begin{eqnarray}
\nonumber
\fl U(2H-1,2H-\mu;s\tau_{\star})&=&\frac{\pi}{\sin(\pi[2H-\mu])}\left[\frac{M(2H-1,
2H-\mu;s\tau_{\star})}{\Gamma(\mu)\Gamma(2H-\mu)}\right.\\
\fl&&\left.-(s\tau_{\star})^{\mu+1-2H}\frac{M(\mu,\mu+2-2H;s\tau_{\star})}{\Gamma(
2H-1)\Gamma(\mu+2-2H)}\right].
\label{mainardi}
\end{eqnarray}
Taking into account the series expansion of the Kummer function ((13.1.2) in
\cite{Abramowitz}) we consider the following two possibilities:

\subsubsection{Weak power-law truncation, $0<\mu<2H-1<1$.}

In this case the second term in (\ref{mainardi}) is dominant at small $s$ and
thus
\begin{equation}
\label{domin}
\fl U(2H-1,2H-\mu;s\tau_{\star})\sim\frac{\pi(s\tau_{\star})^{1+\mu-2H}}{\sin(\pi[
2H-\mu-1])\Gamma(2H-1)\Gamma(\mu+2-2H)}.
\end{equation}
Plugging this leading behaviour into expressions (\ref{msd power-law truncation gle})
and (\ref{abbrevg}) and using the Tauberian theorem, after few transformations we
obtain the long time behaviour of the MSD,
\begin{equation}
\label{msdlim}
\langle x^2(t)\rangle\sim\frac{\Gamma(2H-1)}{\Gamma(2H-\mu-1)\Gamma(\mu+3-2H)}
\frac{2k_BT}{m\Gamma_H\tau_{\star}^{\mu}}t^{\mu+2-2H}.
\end{equation}
Note that in the limit $\mu\to0$ expression (\ref{msdlim}) reduces to the
untruncated formula (\ref{msd fractional noise gle}). Thus, since we observe
the inequality $0<\mu+2-2H<1$ in the present case of a weak power-law truncation,
the dynamics is still subdiffusive, however, with an anomalous diffusion exponent
larger than the value $2-2H$ in the short time limit.

\subsubsection{Strong power-law truncation, $\mu>2H-1>0$.}

In this case the first term in the square brackets in equation (\ref{mainardi})
becomes dominant at small $s$ and $U(2H-1,2H-\mu;s\tau_{\star})\sim\Gamma(\mu
+1-2H)/\Gamma(\mu)$, where we made us of the reflection formula for the Gamma
function. From results (\ref{msd power-law truncation gle}) and (\ref{abbrevg})
by use of the Tauberian theorem we obtain
\begin{equation}
\label{pllong}
\langle x^2(t)\rangle\sim\frac{\Gamma(\mu)}{\Gamma(\mu+1-2H)}\frac{2k_BT}{
m\Gamma_H\tau_{\star}^{2H-1}}t,
\end{equation}
valid for $t\gg\tau_{\star}$. As expected, we find the desired crossover to the
normal Brownian scaling of the MSD. Note that this result is in full accordance
with equation (\ref{msdlong}). Indeed, from expression (\ref{powtrunc}) we get
(see 2.2.5.24 \cite{prudnikov1})
\begin{equation}
\label{newint}
\int_0^{\infty}\gamma_H(\tau)d\tau=\frac{\Gamma(\mu+1-2H)}{\Gamma(\mu)}\Gamma_H
\tau_{\star}^{2H-1}.
\end{equation}
After plugging expression (\ref{newint}) into (\ref{msdlong}) we arrive at
result (\ref{pllong}). Note also that the condition of a strong power-law
truncation is equivalent to the condition that integral (\ref{newint}) converges.

In the borderline case with $0<\mu=2H-1<1$ we use 13.5.9 in \cite{Abramowitz}
and find $U(2H-1,1;s\tau_{\star})\sim-\ln(s\tau_{\star})/\Gamma(2H-1)$. With the
use of the Tauberian theorem equations (\ref{msd power-law truncation gle}) and
(\ref{abbrevg}) yield
\begin{equation}
\label{border1}
\langle x^2(t)\rangle\sim\Gamma(2H-1)\frac{2k_BT}{m\Gamma_H\tau_{\star}^{2H-1}}
\frac{t}{\ln(t/\tau_{\star})}
\end{equation}
at $t\gg\tau_{\star}$. Thus, in this borderline situation between the cases of
weak truncation (leading to increased subdiffusion at long times) and strong
truncation (normal long time diffusion) we observe a logarithmic correlation to
normal diffusion.

\begin{figure}
\centering
\includegraphics[width=10cm]{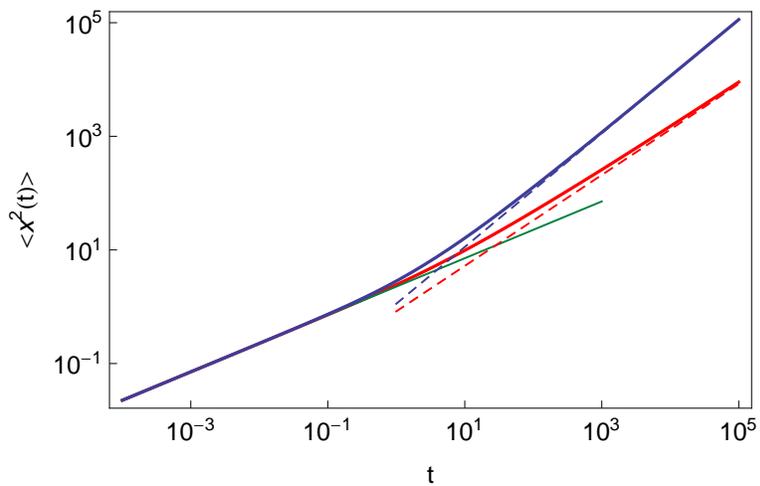}
\caption{MSD (\ref{msd power-law truncation gle}) for power-law truncation with
$H=3/4$, $k_BT/[m\Gamma_H]=1$, and $\tau_{\star}=1$. The red solid
line corresponds to weak power-law truncation with $\mu=0.3$, the blue solid line
to strong truncation with $\mu=1$. The asymptotics (\ref{msdlim}) and (\ref{pllong})
are shown by red and blue dashed lines, respectively. The thin green solid line
corresponds to the MSD (\ref{msd fractional noise gle}) for the untruncated case.}
\label{fgr:GLEpowertrunc}
\end{figure}

Figure \ref{fgr:GLEpowertrunc} shows the crossover dynamics for power-law tempering
for the two possible cases: for weak power-law truncation with $\mu=0.3$ we observe
the predicted crossover from slower to faster subdiffusion, while in the case of
strong power-law truncation the subdiffusive dynamics crosses over to normal
diffusion.

\begin{figure}
\centering
\includegraphics[width=10cm]{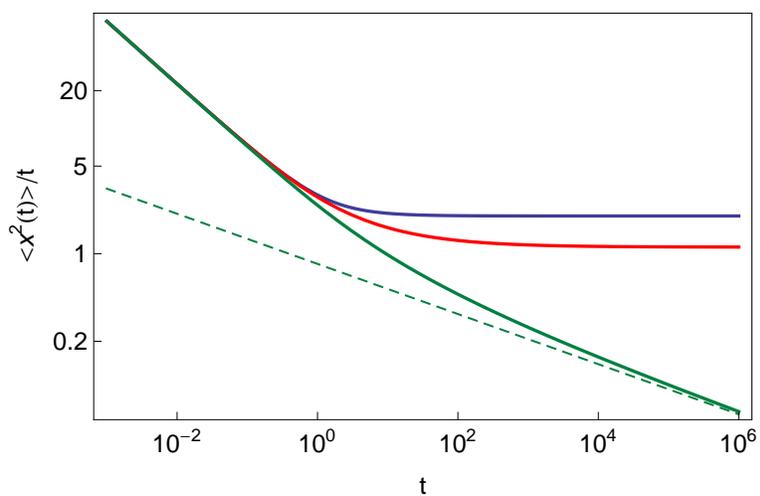}
\caption{Comparison of the ratio $\langle x^2\rangle/t$ for different truncation
modes of the power-law noise in the generalised Langevin equation
(\ref{GLE_overdamped}). Parameters: $H=3/4$, $k_BT/[m\Gamma_H]=1$,
and $\tau_{\star}=1$. From top to bottom the blue line represents the exponential
truncation, equation (\ref{msd exp truncation gle}), the red line the strong
power-law truncation, equation (\ref{msd power-law truncation gle}) with $\mu=1$,
and the green line the weak power-law truncation, equation
(\ref{msd power-law truncation gle}) with $\mu=0.3$. The asymptotics
(\ref{msdlim}) is shown by dashed green line.}
\label{fgr:GLEexpVSpower}
\end{figure}

Figure \ref{fgr:GLEexpVSpower} shows a direct comparison between the cases of
exponential and power-law truncation. As expected, the crossover is faster for
the exponential tempering, and thus the resulting amplitude in this case exceeds
the amplitude for the power-law tempering. Note that the latter observation
contrasts the case of the truncated fractional Brownian motion in figure
\ref{fgr:LEexpVSpower}, for which the amplitude of the power-law tempering is
higher.

\subsubsection{Velocity autocorrelation function.}

To gain some insight into the correlation behaviour we use equation (\ref{auxcorr})
with $\tilde{\gamma}_H(s)$ from equation (\ref{tricomi}). Taking the inverse
Laplace transformation over $s_1$ and $s_2$ in the same way as above we obtain
the position autocorrelation function
\begin{equation}
\langle x(t_1)x(t_2)\rangle=\frac{k_BT}{m\Gamma_H\tau_{\star}^{2H-1}}\Big(g(t_1)+g(t_2)
-g(|t_2-t_1|)\Big),
\end{equation}
where $g(t)$ is given by relation (\ref{abbrevg}). From here the velocity
autocorrelation function is obtained as
\begin{equation}
\label{vsquare}
\langle v^2\rangle_{\tau}=\frac{k_BT}{m\Gamma_H\tau_{\star}^{2H-1}}\frac{d^2}{
d\tau^2}g(\tau)
\end{equation}
with $\tau>0$. We first note that expression (\ref{vsquare}) along with (\ref{abbrevg})
may suggest that the Tauberian theorem may be directly applied to the expression
$U^{-1}(2H-1,2H-\mu;s\tau_{\star})$ in order to calculate the asymptotic behaviour
of the velocity autocorrelation function $\langle v^2\rangle_{\tau}$. However, for
short times corresponding to $s\tau_{\star}\gg1$ the function $U^{-1}\sim(s\tau_{
\star})^{2H-1}$, and since $1/2<H<1$, the Tauberian theorem does not apply as
$2H-1$ is positive. Instead, we should first obtain the asymptotic of $g(\tau)$
at short times $\tau\ll\tau_{\star}$ by use of the Tauberian theorem, and only
then differentiate twice to get the asymptotic of the velocity autocorrelation
function. This way we arrive at expression (\ref{vcorr}). At long times $\tau\gg
\tau_{\star}$ we again consider the cases of weak and strong power-law truncations
separately.

For the weak power-law truncation with $0<\mu<2H-1<1$ the situation is similar
to the short time limit above. Indeed, $U^{-1}\sim(s\tau_{\star})^{2H-1-\mu}$,
see result (\ref{domin}), and the Tauberian theorem does not apply. Instead we
first plug relation (\ref{domin}) into expression (\ref{abbrevg}) and then
apply the Tauberian theorem. Following relation (\ref{vsquare}) we then find
\begin{equation}
\label{vsquare1}
\langle v^2\rangle_{\tau}\sim-C\frac{k_BT}{m\Gamma_H\tau_{\star}^{\mu}}\frac{1}{
\tau^{2H-\mu}},
\end{equation}
where $C=(2H-\mu-1)\pi^{-1}\sin(\pi[2H-\mu-1])\Gamma(2H-1)$ is a positive constant.
Note that for weak power-law truncation we have $1<2H-\mu<2$, and in the limit
$\mu\to0$ expression (\ref{vsquare1}) reduces to the velocity autocorrelation
function (\ref{vcorr}) in absence of truncation. From comparison of result
(\ref{vsquare1}) with (\ref{vcorr}) we see that the autocorrelation function
in the truncated case decays slower than in the untruncated case. This may
appear counter-intuitive, however, it is in agreement with the antipersistent
character of the fractional Langevin equation model in which the MSD scales like
$\simeq t^{2-2H}$ and the velocity autocorrelation function at long times scales
as $\simeq-\tau^{-2H}$ for $1/2<H<1$. This means that a steeper decay of the
velocity autocorrelation function corresponds to a more subdiffusive regime.
In other words, when $H$ is closer to $1/2$ (the subdiffusive regime is closer
to normal diffusion) then the decay of the autocorrelation function is slower.
To see this better consider the effective Hurst index $H_{\mathrm{eff}}=H-\mu/2$.
Then, for weak power-law truncation the MSD scales like $\simeq t^{2-2H_{\mathrm{
eff}}}$ with $1/2<H_{\mathrm{eff}}<H<1$, and the velocity autocorrelation function
decays as $\simeq-\tau^-{2H_{\mathrm{eff}}}$. Thus, in the truncated case the
diffusion becomes closer to normal, as it should be, while the velocity
autocorrelation function decays slower than in the untruncated case, fully
consistent with the antipersistent fractional Langevin equation model.

Now let us turn to the case of strong power-law truncation with $\mu>2H-1>0$
in which for simplicity we assume that $\mu+1-2H\neq n$ where $n\in\mathbb{N}$
is a positive integer. We are interested in the exponent of the power-law decay
of the velocity autocorrelation function. Then expression (\ref{mainardi}) yields
$U(2H-1,2H-\mu;s\tau_{\star})\sim a_0+a_1s+a_2s^2+\ldots a_ks^k+a_{\mu}s^{\mu+1-2H}
+a_{k+1}s^{k+1}+\ldots$, where $a_i$ with $i=0,1,2,\ldots$ are constants that can
be easily found from expansion 13.1.2 in \cite{Abramowitz} for the first Kummer
function in the square brackets of expression (\ref{mainardi}) and $k=[\mu+1-2H]$
denotes the integer part of the corresponding argument in the Landau bracket
$[\cdot]$. Then $U^{-1}(2H-1,2H-\mu;s\tau_{\star})\sim b_0+b_1s+\ldots+b_ks^k+
b_{\mu}s^{\mu+1-2H}+\ldots$ where the $b_i$ with $i=0,1,2,\ldots$ are again
constant factors. From here and with equations (\ref{abbrevg}) and (\ref{vsquare})
we find after application of the Tauberian theorem and subsequent double
differentiation
\begin{equation}
\label{vsquare2}
\langle v^2\rangle_{\tau}\sim-C\frac{k_BT}{m\Gamma_H\tau_{
\star}^{2H-1}}\frac{1}{\tau^{\mu+2-2H}},
\end{equation}
where $C$ is a positive constant. Note that in the borderline case $1>\mu=2H-1
>0$ both expressions (\ref{vsquare1}) and (\ref{vsquare2}) tend to the same
limit resulting in the logarithmic correction to normal diffusion in expression
(\ref{border1}).

\begin{figure}
\centering
\includegraphics[width=10cm]{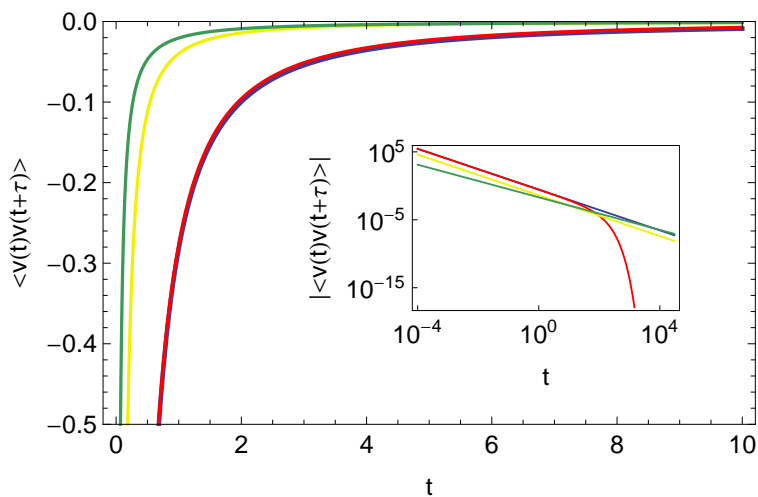}
\caption{Comparison of the velocity autocorrelation functions for the untruncated
case, equation (\ref{vcorr}) (blue line), with exponential truncation, equation
(\ref{vacf exp trunc}) (red line), and with strong power-law truncation, equation
(\ref{vsquare2}) where $\mu=1$ (yellow line), as well as with weak power-law
truncation, equation (\ref{vsquare1}) where $\mu=0.3$ (green line).
Parameters: $H=3/4$, $k_BT/[m\Gamma_H]=1$, and $\tau_{\star}=50$.}
\label{fgr:fig_cv}
\end{figure}

A graphical representation of the velocity autocorrelation function
(\ref{vcorr}), (\ref{vacf exp trunc}) and (\ref{vsquare2}) is shown in figure
\ref{fgr:fig_cv}.

\subsection{Application to lipid molecule dynamics in lipid bilayer membranes}
\label{sec_exp}

We here demonstrate the usefulness of our tempered fractional Gaussian
noise approach to a concrete physical system. The data we have in mind are
from all-atom Molecular Dynamics simulations of lipid bilayer membranes
\cite{ilpo}. In their simplest form, these are double layered leaves made
up of relatively short amphiphilic polymers called lipids. Immersed in
water the double layer arrangement prevents the exposure of the hydrophobic
tail groups to the ambient water, while the hydrophilic head groups are in
contact with the water. At room temperature the lipid bilayer assumes a quite
disordered liquid structure \cite{ilpo}. In this lipid matrix, comparatively
large membrane proteins may be additionally embedded \cite{ilpo}. Natural
biological membranes are composed of lipids of many different chemistries, and
they are crowded with membrane proteins. Supercomputing studies have the task
to reveal the dynamics of both proteins and lipids in such protein-decorated
bilayer systems. This thermally driven diffusion of the constituents influence
biological properties of the bilayer, such as diffusion limited aggregation,
domain formation, or the membrane penetration by nanoparticles \cite{ilpo}.

\begin{figure}
\centering
\includegraphics[width=12cm]{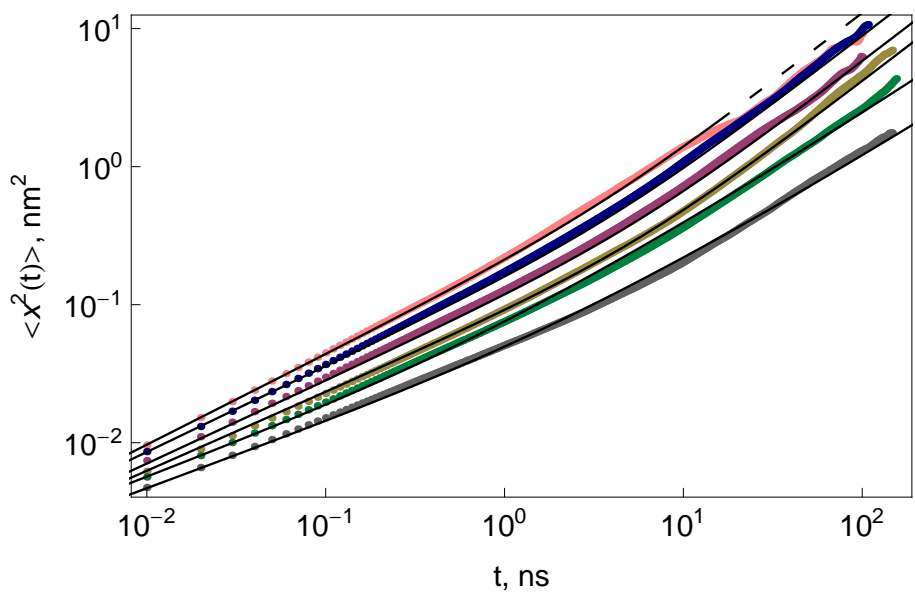}
\caption{MSD of the motion of lipid molecules in a lipid bilayer model membrane,
at room temperature in the liquid disordered and liquid ordered phases (symbols)
\cite{jae_membrane}. The crossover from subdiffusion to normal diffusion or
increased subdiffusion at around 10 nsec is distinct. Data courtesy Matti
Javanainen, University of Helsinki.
The black solid lines provide a fit with equations (\ref{msd exp truncation gle})
and (\ref{msd power-law truncation gle}) resulting from our generalised Langevin
equation model with exponentially and power-law truncated noise, respectively.
The parameters are presented in Table \ref{tab}, see also discussion in the text.}
\label{fgr:MSDexp2}
\end{figure}

Figure \ref{fgr:MSDexp2} depicts the simulations results in a chemically uniform,
liquid disordered lipid bilayer membrane as well as in the liquid ordered state in
the presence of cholesterol molecules---the system is specified in detail in
\cite{jae_membrane}. The motion of the lipids is Gaussian for all cases and best
described as viscoelastic diffusion governed by the generalised Langevin equation
(\ref{GLE_overdamped}) fuelled by power-law noise
\cite{jae_membrane,jae_prx,kneller}.\footnote{Note that the Gaussian character is
lost and intermittent diffusivity dynamics emerge in highly crowded membranes
\cite{jae_prx}, a phenomenon that can be understood
in terms of a superstatistical approach \cite{beck} or within a fluctuating diffusivity
picture \cite{diffdiff,diffdiff1}.} As can be seen in figure \ref{fgr:MSDexp2} the MSD
of the liquid disordered lipid systems exhibits a clear crossover from subdiffusion to
normal diffusion at roughly 10 nsec, the typical crossover time scale discussed in
literature, at which two nearest neighbour lipid molecules exchange their mutual
positions and thus decorrelate their motion \cite{jae_membrane,ilpo,membrane_rev}.
For the liquid ordered cases, one lipid chemistry also shows a subdiffusive-normal
crossover, while the two other lipid chemistries lead to a crossover from slower to
faster subdiffusion \cite{jae_membrane}. From fit of the parameters (see the summary
in table \ref{tab}) to the data we observe an excellent agreement with the short and
long time scaling regimes and, remarkably, the model fully describes the crossover
behaviours without further tuning for both liquid disordered and ordered situations.
We note that subdiffusive-diffusive crossovers are also observed for protein-crowded
membranes \cite{jae_prx,ilpo_faraday,matti_jacs}.

\begin{table}
\hspace*{-1.4cm}
\begin{tabular}{|l|l|l|l|l|l|l|l|l|}
\hline
 & $H$ & $\mu$ & $\tau_{\star}$ & $k_BT/[m\Gamma_H]$ & $\alpha_{\mathrm{short}}$
& $K_{\alpha_{\mathrm{short}}}$ & $\alpha_{\mathrm{long}}$ & $K_{\alpha_{\mathrm{
long}}}$\\
& & & $[\mathrm{nsec}]$ & $[\mathrm{nm}^2/\mathrm{nsec}^{2-2H}]$ & & $[\mathrm{nm}^2/
\mathrm{nsec}^{2-2H}]$ & & $[\mathrm{nm}^2/
\mathrm{nsec}^{\alpha_{\mathrm{long}}}]$ \\\hline\hline
DSPC (purple) & 0.70 & $-$ & 4.0 & 0.050 & 0.60 & 0.034 & 1.0 & 0.029\\\hline
SOPC (pink)   & 0.67 & $-$ & 2.5 & 0.88  & 0.66 & 0.064 & 1.0 & 0.064\\\hline
DOPC (blue)   & 0.69 & $-$ & 3.0 & 0.067 & 0.62 & 0.046 & 1.0 & 0.044\\\hline
\textbf{DSPC} (grey) & 0.76 & 0.41 & 0.60 & 0.019 & 0.48 & 0.010 & 0.89 &
0.0035\\\hline
\textbf{SOPC} (green) & 0.75 & 0.44 & 0.22 & 0.025 & 0.50 & 0.014 & 0.94 &
0.0026\\\hline
\textbf{DOPC} (brown) & 0.72 & $-$ & 4.3 & 0.038 & 0.57 & 0.024 & 1.0 & 0.021\\\hline
\end{tabular}
\caption{Fit parameters for the model membrane simulations data shown in
figure \ref{fgr:MSDexp2}. The colours mentioned in the first column correspond
to the colour coding in figure \ref{fgr:MSDexp2}.}
\label{tab}
\end{table}

We note that from equation (\ref{msd fractional noise gle}) and the effective
diffusion coefficient
\begin{equation}
\label{diffeff}
K^*_{\alpha}(t)=\frac{1}{2}\frac{d}{dt}\left\langle x^{2}(t)\right\rangle.
\end{equation}
we find the short time limiting behaviour
\begin{equation}
K^*_{\alpha}(t)=K^*_{\alpha,\mathrm{short}}t^{1-2H}
\end{equation}
with
\begin{equation}
K^*_{\alpha,\mathrm{short}}=\frac{k_{B}T}{m\Gamma_H}\frac{1}{\Gamma(2-2H)}.
\end{equation}
For the long time limit, from equation (\ref{msd exp truncation
gle long}), it follows that
\begin{equation}
K^*_{\alpha,\mathrm{long}}=\frac{k_{B}T}{m\Gamma_H}\frac{1}{\tau_{\star}^{2H-1}}
\end{equation}
for the exponential tempering, whereas the cases of DSPC and SOPC lipid
chemistries the long time limit in the weak power-law truncation case is given
by
\begin{equation}
K^*_{\alpha,\mathrm{long}}=\frac{\Gamma(2H-1)}{\Gamma(2H-\mu+1)\Gamma(\mu+2-2H)}
\frac{k_BT}{m\Gamma_H\tau_{\star}^{\mu}}
\end{equation} 
The fit values given in table \ref{tab} are in very good agreement with those
obtained in the simulations study \cite{jae_membrane}. We note, however, that
for the weak power-law tempering model fit the crossover time is somewhat
underestimated.

\section{Direct tempering of Mandelbrot's fractional Brownian motion}
\label{sec_tfbm}

So far we introduced the tempering on the level of the noise $\xi(t)$, which
drives the position co-ordinate $x(t)$. Another way to introduce the
crossover from anomalous to normal diffusion is to consider a truncation of
the power-law correlations directly in the original definition of fractional
Brownian motion according
to Mandelbrot and van Ness \cite{mandelbrot}. Such a formulation was recently
proposed by Meerschaert and Sabzikar \cite{Meerschaert20132269}. Here we analyse
this model and demonstrate that it leads to a very different behaviour of the MSD
than the previous tempered fractional models. A formal mathematical analysis of
this model was provided very recently in \cite{Deng_tfle}. We here recall some
of their results for the convenience of the reader and present clear physical
arguments for the seemingly paradoxical behaviour of this model. In particular
we come up with a comparison to a fractional Ornstein-Uhlenbeck scenario.

\subsection{Meerschaert and Sabzikar direct tempering model}

Meerschaert and Sabzikar defined this extension of fractional Brownian motion
by applying an exponential truncating in Mandelbrot's definition
\cite{mandelbrot,Meerschaert20132269},\footnote{Note that in this section we use
dimensionless units in order not to obfuscate the discussion.}
\begin{eqnarray}
\nonumber
B_{H,\lambda}(t)&=&\int_{-\infty}^0\left[e^{-\lambda(t-t')}(t-t')^{H-\frac{1}{2}}
-e^{-\lambda(-t')}(-t')^{H-\frac{1}{2}} \right]B'(t')dt'\\
&&+\int_0^t\left[e^{-\lambda(t-t')}(t-t')^{H-\frac{1}{2}}\right]B'(t')dt',
\label{eq:tfBm_def} 
\end{eqnarray}
where $H,\lambda,t>0$. $B'(t)$ is white Gaussian noise of $\delta$-covariance
$\left\langle B'(t_1)B'(t_2)\right\rangle=\sigma^2\delta(t_1-t_2)$ and zero mean.
The parameter $\lambda$ stands for the truncation parameter, and classical
fractional Brownian motion is
then obtained in the limiting case $\lambda\to0$ when $H\in(0,1)$. It should be
noted that the prefactor $1/\Gamma(H+1/2)$ in Mandelbrot's original definition is
dropped here in line with the procedure of \cite{Meerschaert20132269}. The MSD
encoded in equation (\ref{eq:tfBm_def}) is (see \ref{appmeer} for the
derivation)
\begin{equation}
\label{eq:tfBm_MSD}
\left\langle B^2_{H,\lambda}(t)\right\rangle=\sigma^2 C_t^2t^{2H},
\end{equation}
where the prefactor is
\begin{equation}
C_t^2= \left[\frac{2\Gamma(2H)}{(2\lambda t)^{2H}}-\frac{2\Gamma(H+1/2)}{\sqrt{
\pi}}\frac{K_H(|\lambda t|)}{(2\lambda t)^H}\right].
\end{equation}
$K_H(z)$ denotes the modified Bessel function of the second kind, which for small
argument $z$ behaves as \cite{Abramowitz}
\begin{equation}
\label{bessexp}
K_H(z)\sim\frac{\Gamma(H)}{2^{1-H}}z^{-H}+\frac{\Gamma(-H)}{2^{1+H}}z^{H}+\frac{
\Gamma(H)}{2^{3-H}(1-H)}z^{2-H}
\end{equation}
while for large $z$ we have $K_H(z)\sim\sqrt{\pi/(2z)}e^{-z}$. The fact that the
prefactor $C_t^2$ is an explicit function of time contrasts the result of standard
fractional Brownian motion, and we will readily see the ensuing consequences.

In the short time limit $t\ll\lambda^{-1}$ expression (\ref{eq:tfBm_MSD}) has the
compound power-law form
\begin{equation}
\label{eq:TFBM_MSD_small_t}
\left\langle B_{H,\lambda}^2(t)\right\rangle\sim\sigma^2\Gamma^2(H+1/2)V_H t^{2H}
+\frac{\sigma^2\Gamma(2H)}{2^{1+2H}(H-1)}\lambda^{2-2H}t^2
\end{equation}
with $V_H=1/[\Gamma(2H+1)\sin(\pi H)]$. Thus, the limit $\lambda\to0$ indeed reduces
to the expression for standard fractional Brownian motion. In the long time limit
$t\gg\lambda^{-1}$ the MSD of this tempered fractional Brownian motion, remarkably,
converges exponentially towards a constant value,
\begin{equation}
\label{eq:TFBM_MSD_limit_long_t}
\left\langle B_{H,\lambda}^2(t)\right\rangle\sim\sigma^2\left(\frac{2\Gamma(2H)}{
(2\lambda)^{2H}}-\frac{2^{1/2-H}\Gamma(H+1/2)}{\lambda^{H+1/2}}t^{H-1/2}e^{-\lambda
t}\right),
\end{equation}
a result which is at first surprising. This point will be discussed and compared
to the fractional Ornstein-Uhlenbeck process below. The functional behaviour of
result (\ref{eq:TFBM_MSD_limit_long_t}) is shown in figure \ref{fig:MSD}. We note
that if we consider the Langevin equation (\ref{LE_overdamped}) in combination
with the directly tempered noise $B'_{H\lambda}(t)$, expression (\ref{eq:tfBm_MSD})
and its limiting behaviours (\ref{eq:TFBM_MSD_small_t}) and
(\ref{eq:TFBM_MSD_limit_long_t})
exactly correspond to the dynamics of the MSD $\langle x^2(t)\rangle$.

As shown in \cite{Deng_tfle} it is possible to define a tempered fractional
Gaussian noise following Mandelbrot and van Ness' smoothening procedure
involving a short time lag $\delta$ (see \ref{sec:dem_smooth_tfGn}). The
autocorrelation function of this tempered fractional Gaussian noise is
given through
\begin{eqnarray}
\nonumber
\fl\left\langle B_{H,\lambda}'(t) B_{H,\lambda}'(t+\tau)\right\rangle&=&
\frac{ \Gamma(H+\frac{1}{2})\sigma^2}{\sqrt{\pi}(2\lambda)^H\delta^2}\Big[
2\tau^HK_H(|\lambda\tau|)-(\tau+\delta)^HK_H(\lambda|\tau+\delta|)\\
\fl&&-|\tau-\delta|^HK_H(\lambda|\tau-\delta|)\Big]. 
\label{eq:cov_tfGn_general_delta} 
\end{eqnarray}
An important feature of the autocorrelation function
(\ref{eq:cov_tfGn_general_delta}) for
tempered fractional Gaussian noise is its antipersistent behaviour over the whole
range $0<H<1$ for any finite $\lambda$, that is, the integral of expression
(\ref{eq:cov_tfGn_general_delta}) over the entire domain of $\tau$ vanishes:
\begin{equation}
\label{zeroint}
\int_0^{\infty}\left\langle B_{H,\lambda}'(t)B_{H,\lambda}'(t+\tau)\right\rangle
d\tau=0.
\end{equation}
This is in sharp contrast to (conventional) fractional Gaussian noise. Indeed,
in the limit $\lambda\to0$ the noise autocorrelation function
(\ref{eq:cov_tfGn_general_delta}) approaches the one of fractional Gaussian noise
\cite{mandelbrot,Meerschaert20132269}, as can be derived by using the small argument
expansion (\ref{bessexp}) of the Bessel function. In this limit $\lambda\to0$ for
any finite $\tau$ the  autocorrelation function (\ref{eq:cov_tfGn_general_delta})
converges to
\begin{eqnarray}
\label{eq:cov__FGN} 
\fl\lim_{\lambda\to0}\left\langle B_{H,\lambda}'(t)B_{H,\lambda}'(t+\tau)\right
\rangle\sim\frac{\Gamma^2(H+\frac{1}{2})\sigma^2V_H}{2
\delta^2}\left[(\tau+\delta)^{2H}+|\tau-\delta|^{2H}-2\tau^{2H}\right]
\end{eqnarray}
and shows negative correlations for $0<H<1/2$ and positive correlations for
$1/2<H<1$, see \ref{sec:E_FGN_int1}.

\begin{figure}
\begin{center}
\includegraphics[width=10cm]{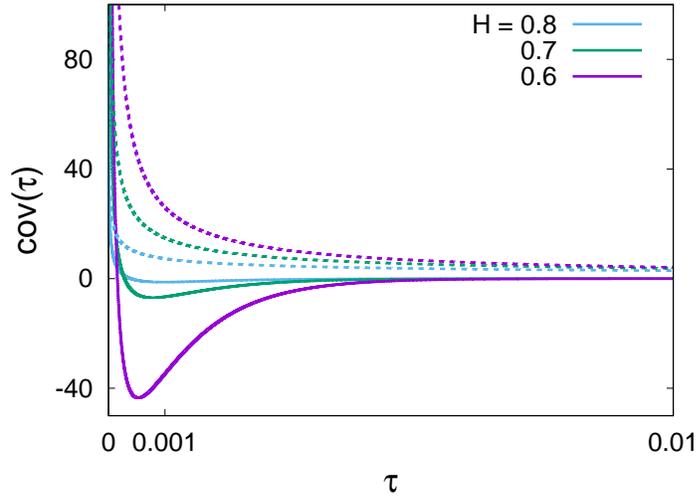} 
\end{center}
\caption{Theoretical results for autocorrelation function,
equations (\ref{eq:cov_tfGn_general_delta}) and (\ref{eq:cov__FGN}), for three
different $H >\frac{1}{2}$ values. The solid lines show the antipersistent
behaviour of autocorrelation function of tempered fractional Gaussian noise, which approaches zero
exponentially; while dashed lines represent the power-law decay of
the autocorrelation function of the fractional Gaussian noise.
Parameters used: $\lambda=10^{3}$,
$\delta=10^{-5}$.}
\label{fig-corr_pers}
\end{figure}

\begin{figure}
\begin{center}
\includegraphics[width=10cm]{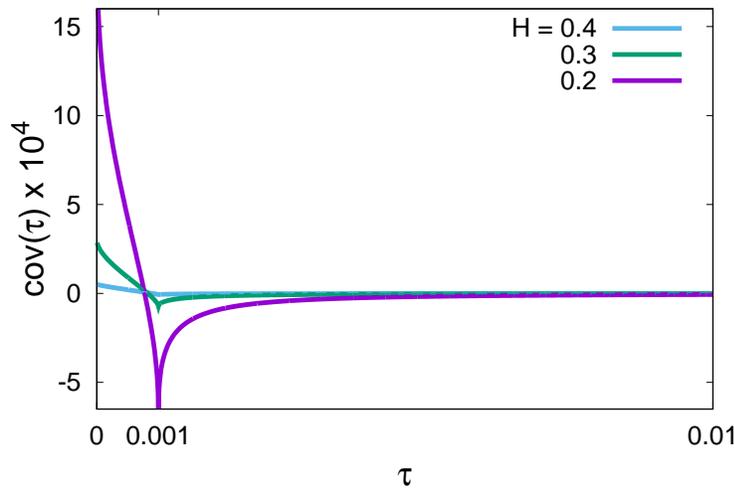} 
\end{center}
\caption{Theoretical results for autocorrelation function,
equations (\ref{eq:cov_tfGn_general_delta}) and (\ref{eq:cov__FGN}), for three
different $H <\frac{1}{2}$ values.  The solid lines show the autocorrelation
function of tempered
fractional Gaussian noise and dashed lines are representation of autocorrelation
function for
fractional Gaussian noise.  There is no significant
difference between the two functions, except around the truncation time,
$\lambda^{-1}$, which is magnified in Fig. (\ref{fig-corr_anti_2}).  Parameters
used: $\lambda=10$, $\delta=10^{-3}$.}
\label{fig-corr_anti}
\end{figure}

\begin{figure}
\begin{center}
\includegraphics[width=10cm]{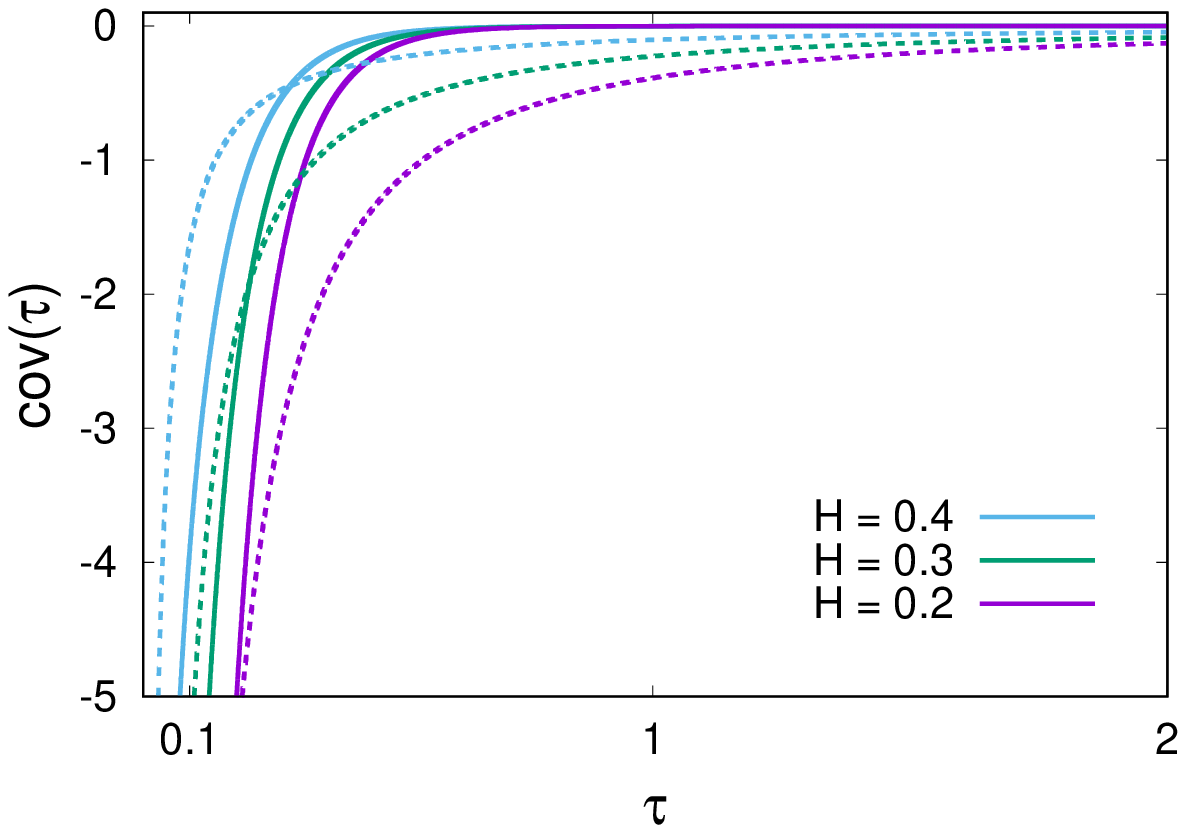} 
\end{center}
\caption{Comparison between the exponentially fast decay of the autocorrelation
function of tempered
fractional Gaussian noise
(solid lines), equation (\ref{E_FGN4}), and the slower power-law decay of its
($\lambda \to 0$) regime, which equivalents to fractional Gaussian noise
(dashed lines),
equation (\ref{ACF_TFGN_t_small}), around the truncation time.  Parameters
used:$\lambda=10$, $\delta=10^{-3}$.}
\label{fig-corr_anti_2}
\end{figure}

The autocorrelation function (\ref{eq:cov_tfGn_general_delta}) and its limit for
$\lambda\to0$ are shown in figures \ref{fig-corr_pers} and \ref{fig-corr_anti} for
different values of the Hurst parameter. While for the tempered process it
is antipersistent for the whole range of $H$, in the limit $\lambda\to0$ we
clearly see the difference between the antipersistent case with the overshoot
to negative values and a slow recovery back to zero. The autocorrelation
function for the persistent case is always positive.

It is easy to show that for $\tau\ll1/\lambda$ and $\delta\to0$ the
autocorrelation function (\ref{eq:cov_tfGn_general_delta}) decays as a power
law, consistent with the behaviour of fractional Gaussian noise,
\begin{eqnarray}
\nonumber
\left\langle B_{H,\lambda}'(t)B_{H,\lambda}'(t+\tau)\right\rangle&\sim&\sigma^2
(2H-1)H\Gamma^2(H+1/2)V_H|\tau|^{2H-2}\\
&&-\frac{\sigma^2\Gamma(2H)\lambda^{2-2H}}{2^{2H+1}(1-H)},
\label{ACF_TFGN_t_small}
\end{eqnarray}
while the asymptotic behaviour at long observation times, $\tau\gg\lambda^{-1} $,
\begin{eqnarray}
\label{E_FGN4} 
\fl\left\langle B_{H,\lambda}'(t)B_{H,\lambda}'(t+\tau)\right\rangle
 \sim\frac{\tau^{H-\frac{1}{2}}e^{-\lambda\tau}\sigma^2}{2^{H-\frac{
1}{2}}\lambda^{H+\frac{1}{2}}\delta^{2}}\left[1-\cosh(\lambda\delta)+\sinh(\lambda
\delta)\frac{(H-\frac{1}{2})\delta}{\tau}\right]
\end{eqnarray}
decays exponentially, in contrast to the non-tempered limit in equation
(\ref{ACF_TFGN_t_small}). This different asymptotic behaviour of tempered
versus non-tempered fractional Gaussian noise around the truncation time,
is shown in figure \ref{fig-corr_anti_2}.

\subsection{Fractional Langevin equation with directly tempered fractional
Gaussian noise}

Considering the internal noise $\xi(t)$ of the system as the tempered fractional
Gaussian noise $B'_{H,\lambda}(t)$ defined above, the overdamped tempered fractional
Langevin equation reads \cite{Deng_tfle}
\begin{eqnarray}
\label{E_FLE1} 
\int_0^t\gamma_H(t-\tau)\frac{dx}{d\tau}d\tau=\xi(t),
\end{eqnarray}
in which $\gamma_H(\tau)=2\left\langle B_{H,\lambda}'(t)B_{H,\lambda}'(t+\tau)
\right\rangle$.
Similar to our derivation above, we obtain the Laplace transform of the MSD
(\ref{lapmsd}) in dimensionless units,
\begin{equation}
\label{tFLE_MSD3} 
\langle\tilde{x}^2(s)\rangle=\frac{2}{s^2\tilde{\gamma}_H(s)},
\end{equation}
in which we have to find the Laplace transformation of the autocorrelation function
(\ref{eq:cov_tfGn_general_delta}).
We assume that $\sigma^2=1$ for simplicity from now on. To proceed, in the second
and third terms we change the variables and split the resulting integrals, 
\begin{eqnarray}
\tilde{\gamma}_H(s)&=&\frac{2\Gamma(H+\frac{1}{2})}{\sqrt{\pi}(2\lambda)^H\delta^2}
\left\{2[1-\cosh(\delta s)]\int_0^{\infty}dte^{-st}t^HK_H(\lambda t)\right.\\
&&+\left.2\int_0^{\delta}dt\sinh(s(\delta-t))t^HK_H(\lambda t)\right\}.
\label{tFLE_MSD5} 
\end{eqnarray}
First, we expand the above functions up to second order in $\delta$. Since in the
second integral $\delta\ll\lambda^{-1}$ and $t<\delta$ the relevant regimes are
$\delta s\ll1$ and $\lambda t\ll1$. Therefore, to second order in $\delta$, $\tilde{
\gamma}_H(s)$ is
\begin{eqnarray}
\nonumber
\tilde{\gamma}_H(s)&\sim&\frac{2\Gamma(H+\frac{1}{2})}{\sqrt{\pi}(2\lambda)^H\delta^2}
\left\{2\frac{-(\delta s)^2}{2}\int_0^{\infty}dte^{-st}t^HK_H(\lambda t)\right.\\
&&\left.+2\int_0^{\delta}dt(s(\delta-t))t^HK_H(\lambda t)\right\}.
\label{tFLE_MSD6} 
\end{eqnarray}
Using expansion (\ref{bessexp})
and keeping terms up to the second order of $\delta$ we find
\begin{equation}
\label{MBS4} 
2\int_0^{\delta}dt(s(\delta-t))t^HK_H(\lambda t)\sim\frac{2^Hs\pi}{\sin(\pi H)
\Gamma(1-H)\lambda^H} \frac{\delta^2}{2}.
\end{equation}
Insertion of this result back to relation (\ref{tFLE_MSD6}) yields
\begin{equation}
\label{tFLE_MSD7} 
\fl\tilde{\gamma}_H(s)\sim\frac{2\Gamma(H+\frac{1}{2})}{\sqrt{\pi}(2\lambda)^H
\delta^2}\left\{\frac{\pi2^{H-1}s\delta^2}{\sin(\pi H)\Gamma(1-H)\lambda^H}-(
\delta s)^2\int_0^{\infty}dte^{-st}t^HK_H(\lambda t)\right\}.
\end{equation}
The integral in (\ref{tFLE_MSD7}) is a Laplace transformation, for which we apply
equation (2.16.6.3) of \cite{Prudnikov}. Hence we find the expression for the
autocorrelation function in Laplace space,
\begin{eqnarray}
\nonumber
\fl\tilde{\gamma}_H(s)&\sim&\frac{2\Gamma(H+\frac{1}{2})}{\sqrt{\pi}(2\lambda)^H
\delta^2}\left\{\frac{\pi2^{H-1}s\delta^2}{\sin(\pi H)\Gamma(1-H)\lambda^H}\right.\\
\fl&&\left.-(\delta s)^2\frac{s^{-1}\lambda^{-H}}{2^{H+1}}\sqrt{\pi}\frac{\Gamma(
2H+1)}{\Gamma(H+3/2)}\,_{2}F_{1}\left(\frac{1}{
2},1;H+\frac{3}{2};1-\frac{\lambda^2}{s^2}\right)\right\}
\label{tFLE_MSD7a} 
\end{eqnarray}
in terms of the hypergeometric function $_2F_1$ \cite{Abramowitz}.

\subsubsection{Short time behaviour of the MSD}

For the regime of short observation times, $\delta\ll t\ll1/\lambda$ we apply
the linear transformations for hypergeometric functions (for more details see
(\ref{sec:MSD_short_t})). Then, with the general definition for hypergeometric
functions up to second order and some simplifications, we find the dominant term
for the autocorrelation function,
\begin{equation}
\label{tFLE_MSD13} 
\tilde{\gamma}_H(s)\sim\frac{2\Gamma^2(H+\frac{1}{2})}{2\sin(\pi H)}s^{1-2H}.
\end{equation}
Substituting this into expression (\ref{tFLE_MSD3}), we see that
\begin{equation}
\label{tFLE_MSD14} 
\langle\tilde{x}^2(s)\rangle\sim\frac{2\sin(\pi H)}{2\Gamma^2(H+
\frac{1}{2})}s^{2H-3}.
\end{equation}
By inverse Laplace transformation we find the asymptotic MSD behaviour in time,
\begin{equation}
\label{tFLE_MSD15} 
\langle x^2(t)\rangle\sim\frac{\sin(\pi H)}{\Gamma^2(H+\frac{1}{2})}\frac{t^{2-
2H}}{\Gamma(3-2H)}.
\end{equation}
This result corresponds to subdiffusion for $1/2<H<1$ in agreement with the
findings in section \ref{sec_gle}. For $0<H<1/2$ the behaviour is superdiffusive.

\subsubsection{Long time behaviour of the MSD}

For the long times regime $t\gg1/\lambda$ or $\lambda/s\gg1$ we go back to
expression (\ref{tFLE_MSD7a}) and use the same method as in the previous
subsection (see also (\ref{sec:MSD_long_t})). It can be seen that the
dominant term is a linear function of $s$,
\begin{equation}
\label{tFLE_MSD17} 
\tilde{\gamma}(s)\sim\frac{\sqrt{\pi}}{2\lambda^{2H}}\frac{2\Gamma(H+\frac{1}{2})}{
\sin(\pi H)\Gamma(1-H)}s.
\end{equation}
Getting back to equation (\ref{tFLE_MSD3}) for the MSD, this yields
\begin{equation}
\label{tFLE_MSD18} 
\langle\tilde{x}^2(s)\rangle\sim\frac{2\lambda^{2H}\sin(\pi H)
\Gamma(1-H)}{2\Gamma(H+\frac{1}{2})\sqrt{\pi}}s^{-3}.
\end{equation}
After inverse Laplace transformation, we obtain
\begin{equation}
\label{tFLE_MSD19} 
\langle x^2(t)\rangle\sim\frac{\sin(\pi H)\Gamma(1-H)\lambda^{2H}}{
\Gamma(H+\frac{1}{2})\sqrt{\pi}}t^2=\frac{\sqrt{\pi}\lambda^{2H}}{
\Gamma(H+\frac{1}{2})\Gamma(H)}t^2.
\end{equation}
Thus, at long times this process converges to ballistic diffusion, as already
observed in \cite{Deng_tfle}.

\begin{figure}
\begin{center}
\includegraphics[width=10cm]{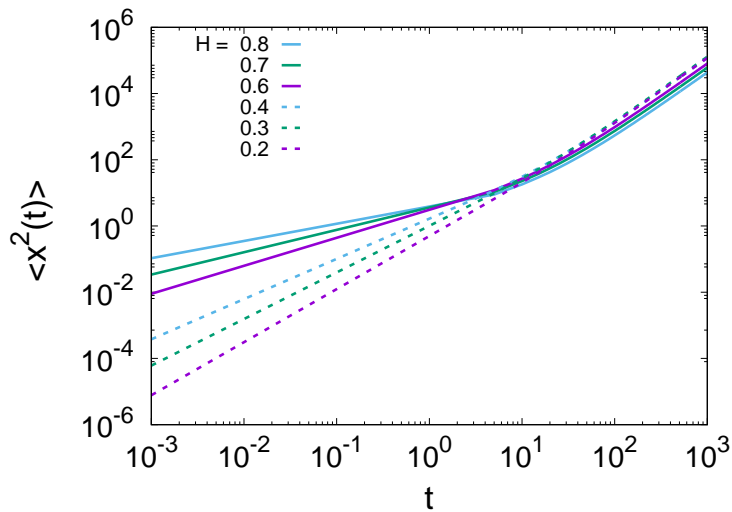} 
\end{center}
\caption{MSD for the tempered Langevin equation (\ref{E_FLE1}), from numerical
Laplace inversion based on result (\ref{tFLE_MSD7a}). We also show the
transition from anomalous diffusion for short time, equation (\ref{tFLE_MSD15}),
to the ballistic regime for long observation times, equation (\ref{tFLE_MSD19}),
is shown for different Hurst exponents and $\lambda=0.1$.}
\label{fig-MSD}
\end{figure}

The general behaviour of the MSD and its crossover from short time power-law
behaviour to long time ballistic motion is shown in figure \ref{fig-MSD} for
different Hurst exponents.

\subsection{Physical discussion of the direct tempering model and
Ornstein-Uhlenbeck with fractional Gaussian noise}
\label{sec_ou}

To come back to the above observed finite limiting value at long times, encoded
in expression (\ref{eq:TFBM_MSD_limit_long_t}), of the MSD in the tempered
fractional Brownian process we briefly study the confined fractional Brownian
motion in an harmonic potential. Experimentally, such a situation arises, for
instance, when particle tracking is performed with an optical tweezers setup
in a viscoelastic environment \cite{norregaard,lene1}. We thus consider the
Ornstein-Uhlenbeck process
\begin{equation}
\label{eq:OU_def}
\frac{dx(t)}{dt}=-\lambda x(t)+B'_H(t),
\end{equation}
for $t>0$ and with $x(0)=0$, where the noise $B'_H(t)$ is again fractional
Gaussian noise. The MSD reads (see \ref{sec:dem_MSD_OU})
\begin{equation}
\label{eq:MSD_fOU_fGn_small_delta}
\langle x^2(t)\rangle=\sigma^2V_Ht^{2H}e^{-\lambda t}\left[1+\frac{\lambda
t}{4H+2}\left(e^{\lambda t}f_H(-\lambda t)-e^{-\lambda t}f_H(\lambda t)\right)
\right],
\end{equation}
where $f_H(x)\equiv M(2H+1;2H+2;x)$ is Kummer's confluent hypergeometric
function. For $t\ll\lambda^{-1}$, the MSD of this fractional Ornstein-Uhlenbeck
process assumes the form
\begin{equation}
\label{eq:MSD_fOU_fGn_small_delta_small_t}
\langle x^2(t)\rangle\sim\sigma^2V_H t^{2H}(1-\lambda t),
\end{equation}
which corresponds to unconfined fractional Brownian motion with a correction
proportional to $\lambda t$. In the long-time limit an exponentially fast
convergence occurs to the stationary limit
\begin{equation}
\label{eq:MSD_fOU_fGn_small_delta_limit_long_t}
\langle x^2(t)\rangle\sim\frac{\sigma^2}{2\sin(\pi H)\lambda^{2H}}.
\end{equation}

\begin{figure}
\centering
\includegraphics[width=16cm]{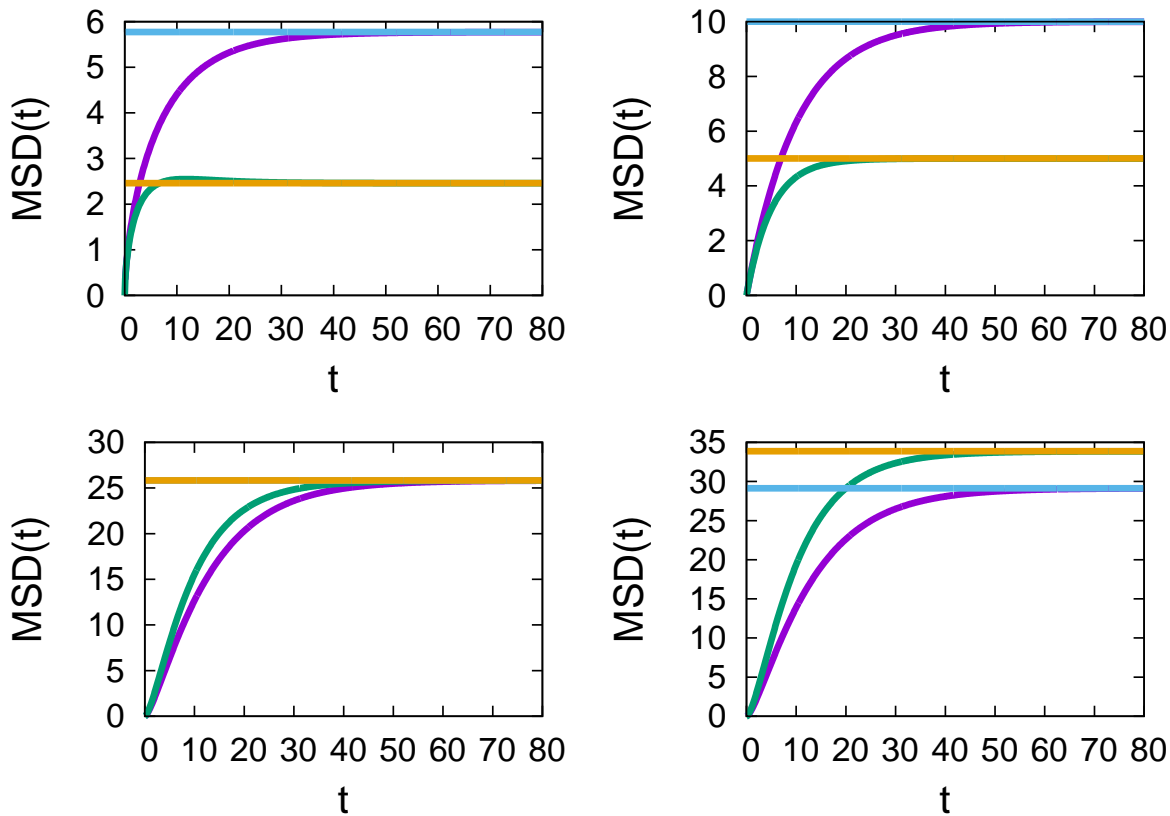}
\caption{MSD of the tempered fractional Brownian motion (equation
(\ref{eq:TFBM_MSD_limit_long_t}), violet line) and the fractional
Ornstein-Uhlenbeck process (equation (\ref{eq:MSD_fOU_fGn_small_delta}), green
line) and their long time plateaus (horizontal lines) for several values of the
Hurst exponent $H$ and the same parameters $\sigma^2=1$ and $\lambda=0.1$. Top
left: $H=0.3$, top right: $H=0.5$, bottom left: $H=0.768149$ (equivalence
of the plateau values), bottom right: $H=0.8$.}
\label{fig:MSD}
\end{figure}

Figure \ref{fig:MSD} compares the MSDs of tempered fractional Brownian motion
and of the fractional Ornstein-Uhlenbeck process. Both of them saturate at
long times, where the plateau value depends on the value of $H$, compare
also \cite{oleksii,jae_pre}. Curiously, the plateau values of both processes
become identical for the Hurst exponent $H=0.768149$.

From the comparison with this fractional Ornstein-Uhlenbeck process we see that
the direct tempering model of Meerschaert and Sabziker actually describes a
confined motion, in contrast to the simple intuition of the tempering in
equation (\ref{eq:tfBm_def}). In that sense it is fundamentally different
from the truncated models considered in the previous sections which show a
crossover between two regimes of steadily increasing MSD.

The effect of direct tempering for the fractional Langevin equation model, a priori
is even more surprising. Namely, as we saw from equations (\ref{tFLE_MSD15}) and
(\ref{tFLE_MSD19}), this model demonstrates a crossover from a short time subdiffusive
to a ballistic regime at long times. Such a behaviour appears counterintuitive.
However, as we show not, it is actually a simple consequence of the two basic
features of the directly tempered internal fractional Gaussian noise
(\ref{E_FLE1}): (i) the integral of its autocorrelation function over the
entire time domain from zero to infinity is identical to zero, see relation
(\ref{zeroint}); (ii) at long times the autocorrelation function exhibits the
exponential decay (\ref{E_FGN4}). To demonstrate that these two conditions
indeed effect the ballistic long time behaviour, consider a toy model for the
noise $\xi(t)$ in the fractional Langevin equation (\ref{E_FLE1}), namely, we
assume the autocorrelation function
\begin{equation}
\label{toyauto}
\langle\xi(t)\xi(t+\tau)\rangle=\gamma_H(\tau)=\delta(\tau)-\lambda e^{-\lambda
\tau}.
\end{equation}
Note that the spectral density of the noise is non-negative and the autocorrelation
function (\ref{toyauto}) obeys conditions (i) and (ii). Now, the Laplace transform
of the autocorrelation function (\ref{toyauto}) reads $\tilde{\gamma}_H(s)=s/(s+
\lambda)$, and with relation (\ref{tFLE_MSD3}) we thus find the MSD
\begin{equation}
\langle\tilde{x}^2(s)\rangle=\frac{2}{s^2\tilde{\gamma}_H(s)}=\frac{2(s+\lambda)}{
s^3},
\end{equation}
in Laplace space. As function of time, this indeed produced the ballistic long
time behaviour $\langle x^2(t)\rangle\sim\lambda t^2$ for $t\gg1/\lambda$.

As we see the direct tempering approach leads to unexpected behaviours. Because
of the stationary limit (\ref{eq:TFBM_MSD_limit_long_t}) the model
by Meerschaert and Sabzikar may be more appropriate for modelling the velocity
process rather than the position of a diffusing particle. Conversely, the
emergence of the ballistic motion (\ref{tFLE_MSD19}) at long times for the directly
tempered fractional Langevin equation may find useful applications for active
systems.

\section{Conclusions}
\label{sec_conc}

In finite systems anomalous diffusion is typically a transient phenomenon,
albeit the crossover time to normal diffusive behaviour may be beyond the
observation window of the experiment or simulations. In those analyses that
explicitly monitor the anomalous-to-normal diffusive crossover, it is desirable
to have a complete quantitative model combining the initial anomalous and the
terminal normal diffusive regimes, instead of a naive fitting
of a non-linear ($\alpha\neq1$) and a linear ($\alpha=1$) power-law for the mean
squared displacement. The explicit
analytical results obtained here provide a two-parameter (exponential cutoff)
or three-parameter (power-law cutoff) model for such crossover dynamics and
thus have the additional advantage of allowing one to extract the crossover
time $\tau_{\star}$ in those cases when the crossover is rather prolonged and
$\tau_{\star}$ otherwise difficult to extract.  Considering systems driven by
Gaussian yet power-law correlated noise we introduced two types of tempering
of these correlations, a hard exponential and a softer power-law truncation. By
plugging this persistent noise into the regular Langevin equation, we
produce a superdiffusive-normal diffusive crossover, as would be observed for
actively moving but eventually decorrelating particle or animals. In contrast,
when we fuel the generalised Langevin equation with this noise, due to the
fluctuation dissipation relation the resulting motion becomes antipersistent,
and the tempering leads to a subdiffusion-normal diffusion crossover. For
the latter case we explicitly showed that the tempered anomalous diffusion
model is very useful for the quantitative description of simulations data
of lipid molecules in a lipid bilayer membrane. Including the shape of the
crossover regime excellent agreement between data and model are observed.

Autocorrelation functions, as studied here, of time series can be directly related
to the distribution of first passage times, that is, the distribution of times between
consecutive zero crossings of the time series \cite{carretero}. More recently, the
first passage time distribution was studied in the presence of crossovers in the
autocorrelation
function of the series \cite{carpena}. In that work the authors demonstrate that the
presence of a crossover in the autocorrelation function is related with a crossover
in the first passage time distribution which is in fact much more complicated to
determine. It will
be interesting to explore such a connection for the crossover behaviour studied
herein.

We also note here that there exist other classes of anomalous diffusion models
such as semi-Markovian continuous time random walks with scale-free waiting time
statistic \cite{scher}, Markovian continuous time random walks with time scale
populations \cite{gianninew}, scaled Brownian motion \cite{sbm}, heterogeneous
diffusion processes \cite{hdp}, generalised grey Brownian motion \cite{diffdiff1,
grey}, or a recent approach using heterogeneous Brownian particle ensembles
\cite{gianninewnew}. The use of either model depends on
the physical situation. The motion fuelled by fractional Gaussian noise
considered here is useful for a large range of systems, in particular,
the motion of submicron tracer particles in living biological cells and
artificially crowded environments, or the motion of membrane constituents in
pure and protein decorated lipid bilayer membranes. Similarly, applications
to stochastic transport in other fields such as sediment transport in earth
science \cite{rina} are conceivable. To identify such type of
motion it is not always sufficient to only look at the MSD of the particle
motion, instead, a range of complementary quantitative measures should be
considered \cite{pccp,yasmine}. To analyse the exact behaviour of these
measures for the tempered motion analysed here, including the statistics of
time averaged observables \cite{pccp,maria}, will be the focus of future work.

\appendix

\section{Spectral densities of truncated Gaussian noise}
\label{appa}

At first we check the positivity of the spectral density of the noise
(\ref{correlation_power-law}). Defining the autocorrelation function
$\langle v^2\rangle_{\tau}$ as symmetric function of the time $\tau$ on the
infinite axis with respect to $\tau=0$, the power spectrum becomes
\begin{eqnarray}
\nonumber
\langle\tilde{v}^2\rangle_{\omega}&=&\int_{-\infty}^{\infty}d\tau\langle v^2\rangle_
{\tau}e^{\imath\omega\tau}=2\int_0^{\infty}d\tau\langle v^2\rangle_{\tau}\cos(
\omega\tau)\\
&=&\frac{2\mathcal{D}_H}{\omega^{2H-1}}\sin\left(\left[H-\frac{1}{2}\right]\pi
\right),
\label{power-law corr spectrum}
\end{eqnarray}
which is positive since $1/2<H<1$.

Let us check that for the exponential tempering
(\ref{correlation_power-law_exp_truncation}) the spectral density is also
positive:
\begin{equation}
\label{spectral density exp truncation}
\langle\tilde{v}^2\rangle_{\omega}=\frac{2\mathcal{D}_H}{\left(\omega^2+\tau_{
\star}^{-2}\right)^{H-1/2}}\cos\left([2H-1]\arctan(\omega\tau_{\star})\right),
\end{equation}
where we made use of 2.5.31.4 \cite{Prudnikov}. This expression
is non-negative since the argument of the cosine function lies between $-\pi/2$
and $+\pi/2$.

Let us now go to the case of power-law tempering, given by expression
(\ref{correlation_power-law_power-law_truncation}). Using 2.5.7.6
of \cite{prudnikov1} we find that
\begin{eqnarray}
\fl
\langle\tilde{v^2}\rangle_{\omega}=\frac{2\mathcal{D}_H\tau_{\star}^{2H-1}}{
\Gamma(2H-1)}\left[\frac{\Gamma(2H-1)\Gamma(\mu-2H+1)\,{_2}F_{3}\left(\frac{2H-1}{
2},H;\frac{1}{2},\frac{2H-\mu}{2},\frac{2H-\mu+1}{2};-\frac{(\omega\tau_{\star})^2}{
4}\right)}{\Gamma(\mu)}\right.\nonumber\\
\fl
\nonumber
+\frac{\Gamma(2H-\mu-1)}{(\omega\tau_{\star})^{2H-\mu-1}}\cos\left(\frac{[2H-\mu
-1]\pi}{2}\right)\\
\times{_2}F_{3}\left(\frac{\mu}{2},\frac{\mu+1}{2};\frac{1}{2},\frac{\mu
-2H+3}{2},\frac{\mu-2H+2}{2};-\frac{(\omega\tau_{\star})^{2}}{4}\right)\nonumber\\
\nonumber
\fl
+\mu\,\Gamma(2H-\mu-2)\sin\left(\frac{[\mu-2H+1]\pi}{2}\right)\\
\times\left.\frac{{_2}F_{
3}\left(\frac{\mu+1}{2},\frac{\mu+2}{2};\frac{3}{2},\frac{\mu-2H+4}{2},\frac{\mu-2H
+3}{2};-\frac{(\omega\tau_{\star})^{2}}{4}\right)}{(\omega\tau_{\star})^{2H-\mu-2}}
\right].
\label{spectral density power-law truncation}
\end{eqnarray}
The positivity of this expression was checked numerically with Mathematica for
various values of the exponent $\mu$.

We note that since ${_p}F_{q}\left((a_p);(b_q);0\right)=1$ \cite{erdelyi}, we have
\begin{equation}
\lim_{\omega\to0}
\langle\tilde{v}^2\rangle_{\omega}=2\Gamma(2-2H)\mathcal{D}_H\tau_{\star}^{2H-1}>0
\end{equation}
for all $\mu$. Moreover, for $\mu=1$ result (\ref{spectral density power-law
truncation}) can be simplified with the use of the following property of the
generalised hypergeometric function (\cite{prudnikov2} 7.2.3.7):
if for $r$ values of $a_p$ there also exist equal them $r$ values of $b_q$, then
\begin{equation}
{_p}F_{q}\left((a_{p-r}),(c_r);(b_{q-r}),(c_r);z\right)={_{p-r}}F_{q-r}\left((a_
{p-r});(b_{q-r});z\right).
\end{equation}

\section{Mittag-Leffler functions and derivation of equation
(\ref{msd exp truncation gle})}
\label{appb}

The three parameter Mittag-Leffler function is defined by \cite{Prabhakar}
\begin{equation}
\label{ML_three}
E_{\alpha,\beta}^{\delta}(z)=\sum_{k=0}^{\infty}\frac{(\delta)_k}{\Gamma(\alpha
k+\beta)}\frac{z^k}{k!},
\end{equation}
where $(\delta)_k=\Gamma(\delta+k)/\Gamma(\delta)$ is the Pochhammer symbol. Its
Laplace transform is given by \cite{Prabhakar}
\begin{equation}
\label{laplaceML}
\mathcal{L}\left[t^{\beta-1}E_{\alpha,\beta}^{\delta}(-\nu t^{\alpha})\right](s)=
\frac{s^{\alpha\delta-\beta}}{\left(s^{\alpha}+\nu\right)^{\delta}},
\end{equation}
where $\mathrm{Re}(s)>|\nu|^{1/\alpha}$.

From definition (\ref{ML_three}) we conclude that the behaviour of the three
parameter Mittag-Leffler function is the stretched exponential \cite{pre2015}
\begin{eqnarray}
E_{\alpha,\beta}^{\delta}(-t^{\alpha})\simeq\frac{1}{\Gamma(\beta)}-\delta\frac{
t^{\alpha}}{\Gamma(\alpha+\beta)}\simeq\frac{1}{\Gamma(\beta)}\exp\left(-\delta
\frac{\Gamma(\beta)}{\Gamma(\alpha+\beta)}t^{\alpha}\right).
\label{ML three small argument}
\end{eqnarray}

Using the series expansion around $z=\infty$ \cite{fcaa2015} (for details see also
\cite{GG})
\begin{equation}
E_{\alpha,\beta}^{\delta}(-z)=\frac{z^{-\delta}}{\Gamma(\delta)}\sum_{k=0}^{\infty}
\frac{\Gamma(\delta+k)}{\Gamma(\beta-\alpha(\delta+n))}\frac{(-z)^{-n}}{n!},
\end{equation}
for $0<\alpha<2$ and $z\rightarrow\infty$, we find that the asymptotic behaviour of
the three parameter Mittag-Leffler function is given by 
\begin{equation}
E_{\alpha,\beta}^{\delta}(-t^{\alpha})\simeq\frac{t^{-\alpha\delta}}{\Gamma(\beta-
\alpha\delta)}, \quad t\rightarrow\infty.
\end{equation} 

The following formula for the derivative of the Mittag-Leffler function follows
directly from definition (\ref{ML_three}) applying term-by-term differentiation,
\begin{equation}
\label{mldiff}
\frac{d}{dt}\left(t^{\beta-1}E_{\alpha,\beta}^{\delta}\left(at^{\alpha}\right)
\right)=t^{\beta-2}E_{\alpha,\beta-1}^{\delta}\left(at^{\alpha}\right).
\end{equation}

From the generalised Langevin equation (\ref{GLE_overdamped}) and the exponentially
truncated friction kernel (\ref{gamexp}) via the Laplace transform method,
we find for the MSD
\begin{equation}
\label{derivationMSD}
\left\langle x^{2}(t)\right\rangle=\frac{2k_{B}T}{m\Gamma_H}\mathcal{L}^{-1}
\left[\frac{s^{-2}}{(s+\tau_{\star}^{-1})^{1-2H}}\right].
\end{equation}
Therefore, from the Laplace transform formula (\ref{laplaceML}), where $\alpha
\to1$, $\delta\to1-2H$, $\alpha\delta-\beta\to-2$, that is, $\beta\to3-2H$, and
$\nu\to\tau_{\star}^{-1}$, we obtain the result (\ref{msd exp truncation gle}).

\section{Derivations for section \ref{sec_tfbm}}
\label{appmeer}

\subsection{Derivation of MSD for tfBm}
\label{sec:dem_MSD_tfBm}

Due to the white Gaussian noise in equation (\ref{eq:tfBm_def})
the MSD of tempered fractional Brownian motion (\ref{eq:tfBm_def}) can be
written as
\begin{eqnarray}
\nonumber
\left\langle B^2_{H,\lambda}(t)\right\rangle&=&\sigma^2\Bigg[\int_0^t e^{-2
\lambda(t-u)}(t-u)^{2H-1}du\\
&&+\int_{-\infty}^0 \left( e^{-\lambda(t-u)}(t-u)^{H-1/2}-e^{\lambda u}(-u)^{
H-1/2} \right)^2 du\Bigg].
\end{eqnarray}
After expanding the square of the second integral and using the appropriate
changes of variable, it becomes
\begin{eqnarray}
\fl\left\langle B^2_{H,\lambda}(t)\right\rangle=\sigma^2 \Big[\int_0^\infty e^{-2
\lambda ts}s^{2H-1}ds-e^{-\lambda t}\int_0^{\infty}e^{-2\lambda ts}(1+s)^{H-1/2}
s^{H-1/2}ds\Big].
\end{eqnarray}
These integrals can be found, for instance, as equations (3.381 4) and (3.383 8)
in \cite{Gradshteyn}. This produces equation (\ref{eq:tfBm_MSD}).

\subsection{Derivation of autocorrelation function of tempered fractional
Gaussian noise}
\label{sec:dem_smooth_tfGn}

In the classical paper by Mandelbrot and van Ness \cite{mandelbrot} a smooth
fractional Brownian motion is defined in terms of the small and positive parameter
$\delta$, through
\begin{equation}
B_H(t;\delta)=\frac{1}{\delta}\int_t^{t+\delta}B_H(u)du.
\end{equation}
Its derivative is known as the fractional Gaussian noise
\begin{equation}
\label{eq:fGn_as_fBm_diff_1}
B_H'(t;\delta)=\frac{1}{\delta}\left[B_H(t+\delta)-B_H(t)\right],
\end{equation}
where we omit the explicit dependence on $\delta$ in the main text.
The autocorrelation function of equation (\ref{eq:fGn_as_fBm_diff_1}) is given in
expression (\ref{eq:cov__FGN}).

The same procedure can be applied to tempered fractional Brownian motion to
define the corresponding continuous fractional noise
\begin{equation}
B'_{H,\lambda}(t;\delta)=\frac{1}{\delta}\left[B_{H,\lambda}(t+\delta)-B_{H,
\lambda}(t)\right].
\end{equation}
With the identity
\begin{equation}
2(a-b)(c-d)=(a-d)^2+(b-c)^2-(a-c)^2-(b-d)^2,
\end{equation}
and the fact that tempered fractional Brownian motion has stationary increments,
and $B_{H,\lambda}(0)=0$, we obtain
\begin{equation}
\fl\Big\langle B_{H,\lambda}'(t;\delta) B_{H,\lambda}'(t+\tau;\delta)\Big\rangle=
\frac{1}{2\delta^2}\Big[\left\langle B_{H,\lambda}^2(\tau-\delta)\right\rangle+
\left\langle B_{H,\lambda}^2(\tau+\delta)\right\rangle-2\left\langle B_{H,\lambda}
^2(\tau)\right\rangle\Big].
\end{equation}
By virtue of relation (\ref{eq:tfBm_MSD}) the autocorrelation function of tempered
fractional Gaussian noise becomes expression (\ref{eq:cov_tfGn_general_delta}).
The autocorrelation function of tempered fractional Gaussian noise
(\ref{eq:cov_tfGn_general_delta}) has a well defined limit when $\delta\lambda\to0$,
\begin{eqnarray}
\nonumber
\Big<B_{H,\lambda}'(t)B_{H,\lambda}'(t+\tau)\Big>&=&\frac{\sigma^2\Gamma(H+1/2)
\lambda^{2-2H}}{2^{H}\sqrt{\pi}}\Big[(\lambda|\tau|)^{H-1} K_{1-H}(\lambda|\tau|)\\
&&-(\lambda|\tau|)^{H} K_{2-H}(\lambda|\tau|)\Big].
\label{eq:cov_tfGn_delta0}
\end{eqnarray}

\subsection{Evaluating the integral over the autocorrelation function of
fractional Gaussian noise}
\label{sec:E_FGN_int1}

Taking the integral over expression (\ref{eq:cov__FGN}) and denoting
\begin{eqnarray}
\label{simplify1}
W_H=\frac{\Gamma^2(H+\frac{1}{2})}{2\Gamma(2H+1)\sin(\pi H)}
\end{eqnarray}
one gets
\begin{eqnarray}
\label{E_FGN_int2}
\nonumber
\fl\mathbb{K}&=&\int_0^{\infty}d\tau\lim_{\lambda\to0}\left\langle B_{H,\lambda}'(t)
B_{H,\lambda}'(t+\tau)\right\rangle\\
\fl&=&\frac{W_H}{\delta^2}\times\lim_{A\to\infty}\left[\int_0^Ad\tau
(\tau+\delta)^{2H}+\int_0^Ad\tau|\tau-\delta|^{2H}-2\int_0^Ad\tau\tau^{2H}\right]
\nonumber\\ 
\fl&=&\frac{W_H}{\delta^2}\times\lim_{A\to\infty}\left[\int_{\delta}^{A+\delta}d
\tau\tau^{2H}+\int_{-\delta}^{A-\delta}d\tau\tau^{2H}-2\int_0^Ad\tau\tau^{2H}\right]
\nonumber\\ 
\fl&=&\frac{W_H}{(2H+1)\delta^2}\times\lim_{A\to\infty}\left[(A+\delta)^{2H+1}-
\delta^{2H+1}+(A-\delta)^{2H+1}-\delta^{2H+1}-2A^{2H+1}\right]\nonumber\\ 
\fl&=&\frac{W_H}{(2H+1)\delta^2}\times\lim_{A\to\infty}\left[A^{2H+1}\left(1+(
2H+1)\frac{\delta}{A}+2H(2H+1)\frac{\delta^2}{2 A^2}\right)-\delta^{2H+1}\right.
\nonumber\\
\fl&&\left.+A^{2H+1}\left(1-(2H+1)\frac{\delta}{A}+2H(2H+1)\frac{\delta^2}{2 A^2}
\right)-\delta^{2H+1}-2A^{2H+1}\right]\nonumber\\ 
\fl&=&\frac{W_H}{(2H+1)\delta^2}\times\lim_{A\to\infty}\left[2H(2H+1)\delta^2A^{
2H-1}\right]\nonumber\\ 
\nonumber
\fl&=&\frac{\Gamma^2(H+\frac{1}{2})}{2\Gamma(2H)\sin(\pi H)}\times\lim_{A\to\infty}
\left[A^{2H-1}\right]\\
\fl&=&\left\{\begin{array}{lr}\infty,&H>\frac{1}{2}\\0,&H<\frac{1}{2}\end{array}
\right..
\end{eqnarray}

\subsection{Tempered fractional Gaussian noise: MSD for short observation times}
\label{sec:MSD_short_t}

For the regime of short observation times, $\delta\ll t\ll\lambda^{-1}$,
we apply the linear transformation 15.3.6 from \cite{Abramowitz} for
hypergeometric functions. In the resulted definition, the argument of the
hypergeometric function is small,
\begin{eqnarray}
\fl\tilde{\gamma}_H(s)&=&\frac{2\Gamma(H+\frac{1}{2})}{\sqrt{\pi}(2\lambda)^H}\left\{
\frac{\pi2^{H-1}s}{\sin(\pi H)\Gamma(1-H)\lambda^H}-s^2\frac{s^{-1}\lambda^{-H}}{
2^{H+1}}\sqrt{\pi}\Gamma\left[\begin{array}{l}1,2H+1\\H+\frac{3}{2}\end{array}
\right]\right.\nonumber\\
\nonumber
\fl&&\times\left[\frac{\Gamma(H+\frac{3}{2})\Gamma(H)}{\Gamma(H+1)\Gamma(H+
\frac{1}{2})}\, _2F_1\left(\frac{1}{2},1;1-H;\frac{\lambda^2}{s^2}\right)\right.\\
\fl&&+\left.\left.\left(\frac{\lambda^2}{s^2}\right)^{H} \frac{\Gamma(H+\frac{3}{2})
\Gamma(-H)}{\Gamma(\frac{1}{2})\Gamma(1)}\,_2F_1\left(H+1,H+\frac{1}{2};H+1;\frac{
\lambda^2}{s^2}\right)\right]\right\}.
\label{tFLE_MSD8} 
\end{eqnarray}
For small arguments we use the general definition of hypergeometric functions,
15.1.1 in \cite{Abramowitz}, up to the second order. Then
\begin{eqnarray}
\nonumber
\fl\tilde{\gamma}_H(s)&=&\frac{2\Gamma(H+\frac{1}{2})}{\sqrt{\pi}(2\lambda)^{H}}
\left\{\frac{\pi2^{H-1}s}{\sin(\pi H)\Gamma(1-H)\lambda^H}-\frac{\sqrt{\pi}s}{
2^{H+1}\lambda^H}\Gamma\left[\begin{array}{l}1,2H+1\\H+\frac{3}{2}\end{array}
\right]\right.\nonumber\\
\fl&&\times\left[\frac{\Gamma\left(H+\frac{3}{2}\right)\Gamma(H)}{\Gamma
(H+1)\Gamma\left(H+\frac{1}{2}\right)}\frac{\Gamma(1-H)}{\Gamma\left(\frac{1}{2}
\right)}\left(\frac{\Gamma(\frac{1}{2})}{\Gamma(1-H)}+\frac{\Gamma(1+\frac{1}{2})
\Gamma(2)}{\Gamma(2-H)}\frac{\lambda^2}{s^2}\right)\right.\nonumber\\
\nonumber
\fl&&+\left(\frac{\lambda^2}{s^2}\right)^H\frac{\Gamma(H+\frac{3}{2})\Gamma(-H)}{
\Gamma(\frac{1}{2})}\frac{\Gamma(H+1)}{\Gamma\left(H+\frac{1}{2}\right)\Gamma(H+1)}
\left(\Gamma\left(H+\frac{1}{2}\right)\right.\\
\fl&&\left.\left.\left.+\Gamma\left(H+\frac{3}{2}\right)\frac{\lambda^2}{s^2}\right)
\right]\right\}.
\label{tFLE_MSD10} 
\end{eqnarray}
Now, we simplify the Gamma functions using the duplication formula 6.1.18 in
\cite{Abramowitz},
\begin{eqnarray}
\fl\tilde{\gamma}_H(s)=\frac{2\Gamma(H+\frac{1}{2})}{\sqrt{\pi}(2\lambda)^H}&\left\{
\frac{\pi2^{H-1}s}{\sin(\pi H)\Gamma(1-H)\lambda^H}-\frac{2^{H-1}\Gamma(H)s}{
\lambda^H}-\frac{2^{H-1}\Gamma(H)s}{\lambda^H2(1-H)}\frac{\lambda^2}{s^2}\right.
\nonumber\\
\fl&\left.-\frac{2^{H-1}s}{\lambda^{H}}\frac{H}{(H+\frac{1}{2})}\frac{\Gamma(H)
\Gamma(-H)\Gamma(H+\frac{3}{2})}{\Gamma(\frac{1}{2})}\left(\frac{\lambda^2}{s^2}
\right)^H\right.\nonumber\\
\fl&\left.-\frac{2^{H-1}s}{\lambda^{H}}\frac{H}{(H+\frac{1}{2})}\frac{\Gamma(H)
\Gamma(-H)\Gamma(H+\frac{3}{2})}{\Gamma(\frac{1}{2})}(H+\frac{1}{2})\left(
\frac{\lambda^2}{s^2}\right)^{H+1}
\label{tFLE_MSD12} 
\right\}
\end{eqnarray}
Using Euler's reflection formula, 
\begin{equation}
\label{ERF} 
\Gamma(z)\Gamma(1-z)=\frac{\pi}{\sin(\pi z)}
\end{equation}
the first two terms cancel each other and it can be seen that the dominant term
in the autocorrelation function scales as $s^{1-2H}$,
\begin{eqnarray}
\nonumber
\fl\tilde{\gamma}_H(s)=\frac{2\Gamma(H+\frac{1}{2})}{\sqrt{\pi}(2\lambda)^H}&\left\{
\frac{2^{H-1}}{\lambda^{H-1}(H+\frac{1}{2})}\frac{\pi}{\sin(\pi H)}\frac{\Gamma(
H+\frac{3}{2})}{\sqrt{\pi}}\left(\frac{\lambda}{s}\right)^{2H-1}\right.\\
\fl&\left.-\frac{2^{H-2}\Gamma(H)}{\lambda^{H-1}(1-H)}\left(\frac{\lambda}{s}\right) 
+\frac{2^{H-1}}{\lambda^{H-1}}\frac{\pi \Gamma(H+\frac{3}{2})}{\sin(\pi H)\sqrt{
\pi}}\left(\frac{\lambda}{s}\right)^{2H+1}\right\}.
\end{eqnarray}

\subsection{Tempered fractional Gaussian noise: MSD for long observation time}
\label{sec:MSD_long_t}

For the regime of long observation time or $\frac{\lambda}{s}\gg1$, we go back
to equation (\ref{tFLE_MSD7a}) and use relation (15.3.8) from \cite{Abramowitz}
for hypergeometric functions with small arguments. Then, by applying the expansion
of hypergeometric functions up to the second order for small argument, $s/\lambda
\ll1$,
\begin{eqnarray}
\fl&&_2F_1\left(\frac{1}{2},1;H+\frac{3}{2};1-\frac{\lambda^2}{s^2}\right)=\left(
\frac{s}{\lambda}\right)\frac{\Gamma(H+\frac{3}{2})\Gamma(\frac{1}{2})}{\Gamma(
1)\Gamma(H+1)}\,_2F_1\left(\frac{1}{2},H+\frac{1}{2};\frac{1}{2};\frac{s^2}{
\lambda^2}\right)\nonumber\\
\fl&&+\left(\frac{s^2}{\lambda^2}\right)\frac{\Gamma(H+\frac{3}{2})\Gamma(-\frac{
1}{2})}{\Gamma(\frac{1}{2})\Gamma(H+\frac{1}{2})}\,_2F_1\left(1,H+1;\frac{3}{2};
\frac{s^2}{\lambda^2}\right)\nonumber\\
\fl&&=\left(\frac{s}{\lambda}\right)\frac{\Gamma(H+\frac{3}{2})\sqrt{\pi}}{\Gamma
(H+1)}\left[\frac{\Gamma(\frac{1}{2})}{\Gamma(\frac{1}{2})\Gamma(H+\frac{1}{2})}
\sum_{k=0}^{\infty}\frac{\Gamma(k+\frac{1}{2})\Gamma(k+H+\frac{1}{2})}{\Gamma(k+
\frac{1}{2})}\frac{1}{k!}\left(\frac{s^2}{\lambda^2}\right)^k\right]\nonumber\\
\fl&&+\left(\frac{s^2}{\lambda^2}\right)(H+\frac{1}{2})(-2)\left[\frac{\Gamma(
\frac{3}{2})}{\Gamma(1)\Gamma(H+1)}\sum_{k=0}^{\infty}\frac{\Gamma(k+1)\Gamma(
k+H+1)}{\Gamma(k+\frac{3}{2})} \frac{1}{k!}\left(\frac{s^2}{\lambda^2}\right)^k
\right]\nonumber\\
\fl&&=\frac{s}{\lambda}\frac{\left(H+\frac{1}{2}\right)\sqrt{\pi}}{\Gamma(H+1)}
\left[\Gamma\left(H+\frac{1}{2}\right)+\Gamma\left(H+\frac{3}{2}\right)\frac{s^2}{
\lambda^2}\right]\nonumber\\
\fl&&-\frac{s^2}{\lambda^2}\frac{\left(H+\frac{1}{2}\right)\sqrt{\pi}}{\Gamma(
H+1)}\left[\frac{\Gamma(H+1)}{\Gamma\left(\frac{3}{2}\right)}+\frac{\Gamma(2)
\Gamma(H+2)}{\Gamma\left(\frac{3}{2}+1\right)}\frac{s^2}{\lambda^2} \right].
\label{hyper_geo6} 
\end{eqnarray}
As a result, the integral in expression (\ref{tFLE_MSD7a}) is approximated as
\begin{eqnarray}
\nonumber
\fl\int_0^{\infty}dte^{-st}t^HK_H(\lambda t)&\sim&s^{-1}2^{H-1}\lambda^{-H}
\left\{\sqrt{\pi}\Gamma(H+\frac{1}{2})\frac{s}{\lambda}\right.\\
\fl&&\hspace*{-1.2cm}
\left.+\sqrt{\pi}\Gamma(H+\frac{3}{2})\frac{s^3}{\lambda^3}-2\Gamma(H+1)
\frac{s^2}{\lambda^2}-\frac{4}{3}(H+1)\Gamma(H+1)\frac{s^4}{\lambda^4}\right\}
\label{hyper_geo7} 
\end{eqnarray}
Applying these approximations, the resulting expression for the autocorrelation
function in the Laplace domain is
\begin{eqnarray}
\tilde{\gamma}(s)=\frac{2\Gamma(H+\frac{1}{2})}{\sqrt{\pi}(2\lambda)^H}&\left\{
\frac{\pi2^{H-1}s}{\sin(\pi H)\Gamma(1-H)\lambda^H}-s^2 s^{-1}2^{H-1}\lambda^{-H}
\right.\nonumber\\
\nonumber
&\times\left[\sqrt{\pi}\Gamma(H+\frac{1}{2})\frac{s}{\lambda}+\sqrt{\pi}\Gamma(H+
\frac{3}{2})\frac{s^3}{\lambda^3}-2\Gamma(H+1)\frac{s^2}{\lambda^2}\right.\\
&\left.\left.-\frac{4}{3}(H+1)\Gamma(H+1)\frac{s^4}{\lambda^4}\right]\right\}.
\label{tFLE_MSD16} 
\end{eqnarray}
It can be seen that the dominant term is a linear function of $s$,
\begin{equation}
\label{tFLE_MSD17_1} 
\tilde{\gamma}(s)=\frac{2\Gamma(H+\frac{1}{2})}{\sqrt{\pi}(2\lambda)^H}\frac{\pi
2^{H-1}s}{\sin(\pi H)\Gamma(1-H)\lambda^H}
\end{equation}

\section{Derivation of the MSD of the fractional Ornstein-Uhlenbeck process}
\label{sec:dem_MSD_OU}

The solution of equation (\ref{eq:OU_def}) for a general noise $\xi(u)$ is
\begin{equation}
x(t)=e^{-\lambda t}\int_0^t e^{\lambda u} \xi(u)\, du,
\end{equation}
so
\begin{equation}
\label{eq:MSD_OU}
\left\langle x^2(t)\right\rangle = e^{-2\lambda t}\int_0^t\int_0^t e^{\lambda
(u_1+u_2)}\langle\xi(u_1)\xi(u_2)\rangle du_1du_2.
\end{equation}
In general, for a noise such that $\langle\xi(u_1)\xi(u_2)\rangle=g(|u_1-u_2|)$,
equation (\ref{eq:MSD_OU}) becomes
\begin{equation}
\left\langle x^2(t)\right\rangle=\frac{1}{\lambda}\left[\int_0^te^{-\lambda\tau}
g(\tau)d\tau-e^{-2\lambda t}\int_0^te^{\lambda\tau} g(\tau)d\tau\right].
\end{equation}
In our case, $\xi(u)=B'_H(u)$. For $H\neq 1/2$, $g(u)=\sigma^2H(2H-1)V_Hu^{2H-2}$
and the MSD can be expressed in terms of the Kummer function $M(a;b;z)$,
\begin{eqnarray}
\nonumber
\left\langle x^2(t)\right\rangle&=&\frac{\sigma^2 H V_Ht^{2H-1}}{
\lambda}\Big[M(2H-1;2H;-\lambda t)\\
&&-e^{-2\lambda t}M(2H-1;2H;\lambda t)\Big].
\label{eq:MSD_OU_from_delta0}
\end{eqnarray}
If $H=1/2$, using $g(u)=\sigma^2\delta(u)$ in equation (\ref{eq:MSD_OU}), we
arrive at
\begin{eqnarray}
\langle x^2(t)\rangle=\frac{\sigma^2}{2\lambda}\left(1-e^{2\lambda t}
\right).
\end{eqnarray}
This result coincides with equation (\ref{eq:MSD_OU_from_delta0})
for $H=1/2$, such that equation (\ref{eq:MSD_OU_from_delta0}) is
valid for all $H\in(0,1)$. Using the properties of the Kummer function
(which in our case reduces to the incomplete gamma function), relation
(\ref{eq:MSD_OU_from_delta0}) is shown to be equivalent to equation
(\ref{eq:MSD_fOU_fGn_small_delta}).

% \section{Spectral function of the tempered fractional velocity process}
% 
% The power spectral density of the tempered fractional velocity process
% defined in section \ref{sec_tfbm} is obtained as the Fourier transform of
% the covariance of the tempered fractional Gaussian noise
% \begin{equation}
% S(\omega)=\int_{-\infty}^{\infty} e^{i\omega\tau}\langle B'_{H,\lambda}(t)
% B'_{H,\lambda}(t+\tau)\rangle d\tau.
% \end{equation}
% With equation (\ref{eq:cov_tfGn_delta0}) we obtain the following result for $H>1/2$,
% \begin{equation}
% \label{eq:S_tfGn}
% S(\omega) =  \frac{\sigma^2 \omega^2}{(\lambda^2+\omega^2)^{H+1/2}}.
% \end{equation}
% Using instead the covariance for $\delta>0$ (Appendix
% \ref{sec:dem_smooth_tfGn}), the result (\ref{eq:S_tfGn}) is retrieved for
% every $H$ if $|\omega|\ll \delta^{-1}$. Figure \ref{fig:tfGn_spectral_density}
% shows the power spectral density for different values of $H$ and the limiting
% case $\lambda\to 0$, where $S(\omega)\sim\omega^{1-2H}$.

% \begin{figure}
% \centering \includegraphics[width=10cm]{fig10.eps}
% \caption{Power spectral density of tempered fractional Gaussian noise
% for $\sigma^2=1$, $\lambda=0.1$
% and $\delta=0.01$. Red and solid line: $H=0.3$, black and dotted: $H=0.5$,
% blue and dash-dotted: $H=0.7$, dashed lines are their respective asymptotic
% behaviours for $\lambda\to 0$.}
% \label{fig:tfGn_spectral_density}
% \end{figure}

\ack

This research is supported by the Basque Government through the BERC
2014-2017 and BERC 2018-2021 programmes and by Spanish Ministry of Economy and
Competitiveness MINECO, BCAM Severo Ochoa excellence accreditation SEV-2013-0323,
and project
MTM2016-76016-R "MIP". TS, AC and RM acknowledge funding from the Deutsche
Forschungsgemeinschaft, project ME 1535/6-1. RM acknowledges support from
Deutsche Forschungsgemeinschaft, project ME 1535/7-1, as well as from the
Foundation for Polish Science (Fundacja na rzecz Nauki Polski) in the
framework of a an Alexander von Humboldt Polish Honorary Research
Fellowship.

\end{document}